\documentclass[aps,prl,onecolumn,showpacs,preprintnumbers]{revtex4}
\usepackage{eurosym}
\usepackage{amsmath}
\usepackage{dcolumn}
\usepackage{bm}
\usepackage{subfigure}
\usepackage{amsfonts}
\usepackage{amssymb}
\usepackage{makeidx}
\usepackage{epsfig}
\usepackage{graphicx}

\begin{document}

\title{Macroscopic irreversibility and decay to kinetic equilibrium\\
of the 1-body PDF for finite hard-sphere systems}
\author{Massimo Tessarotto}
\affiliation{Department of Mathematics and Geosciences, University of Trieste, Via
Valerio 12, 34127 Trieste, Italy}
\affiliation{Institute of Physics, Faculty of Philosophy and Science, Silesian University
in Opava, Bezru\v{c}ovo n\'{a}m.13, CZ-74601 Opava, Czech Republic}
\author{Claudio Cremaschini}
\affiliation{Institute of Physics and Research Center for Theoretical Physics and
Astrophysics, Faculty of Philosophy and Science, Silesian University in
Opava, Bezru\v{c}ovo n\'{a}m.13, CZ-74601 Opava, Czech Republic}
\date{\today }

\begin{abstract}
The conditions for the occurrence of the so-called macroscopic
irreversibility property and the related phenomenon of decay to kinetic
equilibrium which may characterize the 1-body probability density function
(PDF) associated with hard-sphere systems are investigated. The problem is
set in the framework of the axiomatic "ab initio" theory of classical
statistical mechanics developed recently and the related establishment of an
exact kinetic equation realized by the Master equation for the same kinetic
PDF. As shown in the paper the task involves the introduction of a suitable
functional of the 1-body PDF, identified here with the \textit{Master
kinetic information}. It is then proved that, provided the same PDF is
prescribed in terms of suitably-smooth, i.e., stochastic, solution of the
Master kinetic equation, the two properties indicated above are indeed
realized.
\end{abstract}

\pacs{05.20.-y, 05.20.Dd, 05.20.Jj, 51.10.+y}
\keywords{kinetic theory, classical statistical mechanics, Boltzmann
equation, H-theorem}
\maketitle

\section{1. Introduction}

The axiomatic theory of Classical Statistical Mechanics (CSM) recently
proposed in a series of papers (see Refs.\cite{noi1,noi2,noi3,noi11}), and
referred to as \emph{ab initio theory of CSM}, provides a self-consistent
pathway to the kinetic theory of hard-sphere systems, as well as in
principle also point particles subject to finite-range interactions \cite%
{Tessarotto1979}. Its theoretical\ basis\ and conditions of validity are
indeed founded on a unique physical realization of the axioms which are set
at the foundations of CSM \cite{noi1,noi2,noi3}, a fact which permits the
treatment of phase-space and kinetic probability density functions (PDF)
which are either realized by stochastic (i.e., ordinary) functions or
distributions such as the $N-$body Dirac Delta (or certainty function \cite%
{Cercignani1969a}).\textbf{\ }This feature is physically-based being due to
the prescription of the collision boundary conditions (CBC, \cite{noi2}),
i.e., the relationship occurring at collision events between incoming and
outgoing multi-body probability density functions PDF.\textbf{\ }The choice
of the appropriate CBC indicated in Ref. \cite{noi2}, denoted as modified
collision boundary condition (MCBC), is actually of crucial importance and
departs from the customary realization/interpretation (of the same axioms)
originally adopted in Boltzmann \cite{Boltzmann1972}, Enskog \cite%
{Enskog,CHAPMA-COWLING} and Grad \cite{Grad} kinetic approaches (for a
review of Grad's kinetic theory based on CSM see also Cercignani \cite%
{Cercignani1975,Cercignani1988}). The same choice implies, in fact, a number
of theoretical and physically-relevant consequences. In particular, it
follows that the new theory:

\begin{itemize}
\item Unlike Enskog theory \cite{noi3} applies also to \emph{finite }$N-$%
\emph{body hard-sphere systems} $S_{N}$\emph{,} namely systems formed by $N$
like smooth hard-spheres of diameter $\sigma $ and mass $m$, in which the
parameters ($N,$ $\sigma ,m$) remain all constant and finite \cite{noi3}. On
the other hand, the same particles are assumed as usual: A) subject to
instantaneous (unary, binary and multiple) elastic collisions which leave
unchanged the particles angular momenta and B) immersed in a stationary
bounded domain $\Omega $ of the Euclidean space $%
\mathbb{R}
^{3}$ with finite canonical measure.

\item Has lead to the discovery \cite{noi3} of an exact kinetic equation
holding globally in time \cite{noi7} (i.e., for all $t\in I\equiv
\mathbb{R}
$) for these systems and denoted as Master kinetic equation (recalled in
Appendix A).\textbf{\ }In other words the Master equation is non-asymptotic
in character with respect to the (finite) parameters ($N,$ $%
\sigma ,m$). In addition the same equation holds under suitable maximal
entropy conditions for the statistical treatment of the so-called
Boltzmann-Sinai classical dynamical system (CDS), which implies that initial
(binary or multi-body) phase-space statistical correlations are assumed
identically vanishing, while at the same time only suitable
uniquely-prescribed configuration-space correlations can arise. As such the
equation generalizes and extends the validity of the Boltzmann and Enskog
kinetic equations and notably applies to arbitrary $1-$body PDFs which can
be realized either in terms of stochastic functions or distributions.

\item Is time-reversal invariant \cite{noi11}, namely the Master kinetic
equation is time-reversal ($\mathcal{TR-}$) symmetric. In other words, the
same equation is invariant with respect to the $\mathcal{TR-}$transformation
\begin{equation}
\left\{
\begin{array}{c}
\tau \equiv t-t_{o}\rightarrow \tau ^{\prime }\equiv t^{\prime
}-t_{o}=-t+t_{o}\equiv -\tau , \\
\mathbf{r}_{1}\rightarrow \mathbf{r}_{1}^{\prime }=\mathbf{r}_{1}, \\
\mathbf{v}_{1}\rightarrow \mathbf{v}_{1}^{\prime }=-\mathbf{v}_{1}.%
\end{array}%
\right.  \label{time-reversal-1}
\end{equation}%
Thus, representing the absolute time as\textbf{\ }$t=\tau +t_{o}$, with $t_{o%
\text{ }}$\ being a prescribed (arbitrary) initial time, it follows that the%
\textbf{\ }$\mathcal{TR-}$transformation leaves invariant the initial time $%
t_{o}$\ and the instantaneous position $\mathbf{r}_{1}=\mathbf{r}%
_{1}(t\equiv t_{o}+\tau )$ of an arbitrary particle, while reversing the
signature (i.e., versus) of its velocity $\mathbf{v}_{1}=\mathbf{v}%
_{1}(t\equiv t_{o}+\tau ).$ Accordingly, thanks to $\mathcal{TR}-$symmetry
the two initial-value problems associated with the Master kinetic equation
in the two cases are related in such a way that, respectively, the initial $%
1-$body PDF at time $t_{o}$, $\rho _{1}^{(N)}(\mathbf{x}_{1},t_{o})\equiv $ $%
\rho _{1(o)}^{(N)}(\mathbf{r}_{1},\mathbf{v}_{1}),$ and the corresponding
time-evolved PDF $\rho _{1}^{(N)}(\mathbf{x}_{1},t)$\ are carried into the $%
\mathcal{TR}-$transformed $1-$body PDFs $\rho _{1(o)\mathcal{TR}}^{(N)}(%
\mathbf{r}_{1}^{\prime },\mathbf{v}_{1}^{\prime })$\ and $\rho _{1\mathcal{TR%
}}^{(N)}(\mathbf{r}_{1}^{\prime },\mathbf{v}_{1}^{\prime },t^{\prime })$%
\textbf{\ }respectively prescribed according to the law%
\begin{eqnarray}
&&\left. \left\{
\begin{array}{c}
\rho _{1(o)}^{(N)}(\mathbf{r}_{1},\mathbf{v}_{1}) \\
\rho _{1}^{(N)}(\mathbf{x}_{1},t)\equiv \rho _{1}^{(N)}(\mathbf{r}_{1},%
\mathbf{v}_{1},t_{o}+\tau )%
\end{array}%
\right. \rightarrow \right.  \notag \\
&&\left\{
\begin{array}{c}
\rho _{1(o)\mathcal{TR}}^{(N)}(\mathbf{r}_{1}^{\prime },\mathbf{v}%
_{1}^{\prime })\equiv \rho _{1(o)}^{(N)}(\mathbf{r}_{1},-\mathbf{v}_{1}) \\
\rho _{1\mathcal{TR}}^{(N)}(\mathbf{r}_{1}^{\prime },\mathbf{v}_{1}^{\prime
},t^{\prime })\equiv \rho _{1}^{(N)}(\mathbf{r}_{1},-\mathbf{v}%
_{1},t_{o}-\tau )%
\end{array}%
\right. .  \label{time-reversal-2}
\end{eqnarray}

\item Conserves the corresponding Boltzmann-Shannon (BS) statistical entropy
\cite{noi11}. This is identified with the phase-space moment
\begin{equation}
S(\rho _{1}^{(N)}(t))\equiv -\int\limits_{\Gamma _{1}}d\mathbf{x}_{1}\rho
_{1}^{(N)}(\mathbf{x}_{1},t)\ln \frac{\rho _{1}^{(N)}(\mathbf{x}_{1},t)}{%
A_{1}},
\end{equation}%
with $\rho _{1}^{(N)}(t)\equiv \rho _{1}^{(N)}(\mathbf{x}_{1},t)$ being an
arbitrary stochastic PDF solution of the Master kinetic equation and $A_{1}$
an arbitrary positive constant such that the initial PDF%
\begin{equation*}
\rho _{1}^{(N)}(t_{o})\equiv \rho _{1}^{(N)}(\mathbf{x}_{1},t_{o})=\rho
_{10}^{(N)}(\mathbf{x}_{1})
\end{equation*}%
is such that the corresponding BS functional $S(\rho _{1}^{(N)}(t_{o}))$\ is
defined. As a consequence it follows that an arbitrary smooth solution of
the Master kinetic equation satisfies the constant H-theorem%
\begin{equation}
\frac{\partial }{\partial t}S(\rho _{1}^{(N)}(t))=0
\end{equation}%
for all $t\in I\equiv
\mathbb{R}
$ (see again related discussion in Ref.\cite{noi11}).
\end{itemize}

Based on the ab initio theory of CSM, in this paper the problem is posed of
the existence of two phenomena which are expected to characterize the
statistical description of finite $N-$body hard-sphere systems and therefore
should lay at the very foundation of CSM and kinetic theory. These are
related to the physical conditions for the possible occurrence of the
so-called \emph{property of macroscopic irreversibility }(PMI) and the
consequent one represented by the \emph{decay to kinetic equilibrium} (DKE)
which characterize the $1-$body (kinetic) PDF in these $N-$body systems,
i.e., when $1-$body-factorized initial conditions are considered for the $N-$%
body Liouville equation \cite{noi3}. The conjecture is that -\ in some sense
in analogy with the ubiquitous character of the ergodicity property which
characterizes hard-sphere systems and hence\textbf{\ }the $S_{N}-$CDS \cite%
{Sinai1970,Sinai1989} - the occurrence of such phenomena should be
independent of the number $N$ of constituent particles of the system and
therefore apply to actual physical systems for which the parameters $(N$,$%
\sigma ,m)$ are obviously all finite.

\subsection{1A - Motivations and background}

Both properties indicated above concern the statistical behavior of an
ensemble $S_{N}$ of like particles which are advanced in time by a suitable $%
N-$body classical dynamical system, identified here with the $S_{N}-$CDS.
Specifically they arise in the context of the kinetic description of the
same CDS, i.e., in terms of the corresponding $1-$body (kinetic) probability
density function (PDF) $\rho _{1}^{(N)}(t)\equiv \rho _{1}^{(N)}(\mathbf{x}%
_{1},t).$ The latter is required to belong to the functional class $\left\{
\rho _{1}^{(N)}(\mathbf{x}_{1},t)\right\} $ of suitably smooth and strictly
positive ordinary functions which are particular solutions of the relevant
kinetic equation.

In fact, PMI should be realized by means of a suitable, but still possibly
non-unique, functional which should be globally defined in the future (i.e.,
for all times $t\geq t_{o}$, being $t_{o}$ the initial time) bounded and
non-negative, and therefore to be identified with the notion of information
measure. Most importantly, however, the same functional, to be referred to
here as \emph{Master kinetic information }(MKI), should also exhibit a
continuously-differentiable and monotonic, i.e., in particular decreasing,
time-dependence.

Regarding, instead, the second property of DKE this concerns\emph{\ }the
asymptotic behavior of the $1-$body PDF $\rho _{1}^{(N)}(t)\equiv \rho
_{1}^{(N)}(\mathbf{x}_{1},t)$ which, accordingly, should be globally defined
and decay for $t\rightarrow +\infty $ to a stationary and spatially-uniform
Maxwellian PDF%
\begin{equation}
\rho _{1M}^{(N)}(\mathbf{v}_{1})=\frac{n_{o}}{\pi ^{3/2}\left(
2T_{o}/m\right) ^{3/2}}\exp \left\{ -\frac{m\left( \mathbf{v}_{1}-\mathbf{V}%
_{o}\right) ^{2}}{2T_{o}}\right\} ,  \label{MAXWELLIAN-PDF}
\end{equation}%
where $\left\{ n_{o}>0,T_{o}>0,\mathbf{V}_{o}\right\} $\emph{\ }are constant
fluid fields.

Both PMI and DKE correspond to physical phenomena which are actually
expected to arise in disparate classical $N-$body systems. The clue for
their realization is represented by the ubiquitous occurrence of kinetic
equilibria and consequently, in principle, also of the corresponding
possible manifestation of macroscopic irreversibility and\emph{\ }decay
processes. Examples of the former ones are in principle easy to be found,
ranging from neutral fluids \cite{ikt1} to collisional/collisionless and
non-relativistic/relativistic gases and plasmas \cite%
{Crema2013a,Crema2013b,Crema2014}.

However, the most notable example is perhaps provided by dilute hard-sphere
systems ("gases") characterized by a large number of particles ($N\equiv
\frac{1}{\varepsilon }\gg 1$) and a small (i.e., infinitesimal) diameter $%
\sigma \sim O(\varepsilon ^{1/2})$ of the same hard-spheres, for which the
Boltzmann equation applies. Indeed\ the Boltzmann equation is actually
specialized to the treatment of dilute hard-sphere systems in the
Boltzmann-Grad limit\ discussed in the Lanford theorem \cite%
{Lanford1974,Lanford1976,Lanford1981} (for a detailed discussion of the
topics in the context of the ab initio-theory see also Ref.\cite{noi11}). In
such a case the $1-$body PDF can be formally obtained by introducing the
Boltzmann-Grad limit operator \cite{noi11}\textbf{\ }%
\begin{equation}
\mathcal{L}_{BG}\equiv \lim_{\substack{ N\equiv \frac{1}{\varepsilon }%
\rightarrow \infty  \\ \sigma \sim O(\varepsilon ^{1/2})}},
\end{equation}%
whereby the limit function $\rho _{1}(\mathbf{x}_{1},t)$ is denoted
\begin{equation}
\rho _{1}(\mathbf{x}_{1},t)=\mathcal{L}_{BG}\rho _{1}^{(N)}(\mathbf{x}%
_{1},t),  \label{BG-limit}
\end{equation}%
and $\rho _{1}(\mathbf{x}_{1},t)$ identifies a particular solution of the
Boltzmann kinetic equation.

Historically, the property of irreversibility indicated above is known to be
related to the Carnot's second Law of Classical Thermodynamics.\textbf{\ }%
More precisely, it is related to the first-principle-proof originally
attempted by Ludwig Boltzmann in 1872 \cite{Boltzmann1972}. Actually it is
generally agreed that both phenomena lie at the very heart of Boltzmann and
Grad kinetic theories \cite{Boltzmann1972,Grad} and the related original
construction of the Boltzmann kinetic equation (1872). In particular, the
goal set by Boltzmann himself in his 1872 paper was the proof of Carnot's
Law providing at the same time also a possible identification of
thermodynamic entropy. This was achieved in terms of what is nowadays known
as Boltzmann-Shannon (BS) statistical entropy, which is identified with the
phase-space moment
\begin{equation}
M_{X_{E}}(\rho _{1}(t))\equiv \int\limits_{\Gamma _{1}}d\mathbf{x}_{1}\rho
_{1}(\mathbf{x}_{1},t)X_{E}(\mathbf{x}_{1},t)=-\int\limits_{\Gamma _{1}}d%
\mathbf{x}_{1}\rho _{1}(\mathbf{x}_{1},t)\ln \frac{\rho _{1}(\mathbf{x}%
_{1},t)}{A_{1}}\equiv S(\rho _{1}(t)).  \label{BS-entropy}
\end{equation}%
Here $X_{E}(\mathbf{x}_{1},t)\equiv -\ln \frac{\rho _{1}(\mathbf{x}_{1},t)}{%
A_{1}},$ $\rho _{1}(\mathbf{x}_{1},t)$ and $A_{1}$ denote respectively the
BS entropy density, an arbitrary particular solution of the Boltzmann
equation for which the same phase-space integral exists and an arbitrary
positive constant. In fact, according to the Boltzmann H-theorem \cite%
{Boltzmann1972} the same functional should satisfy the \emph{entropic
inequality}
\begin{equation}
\frac{\partial }{\partial t}S(\rho _{1}(t))\geq 0,
\label{Entropic enequality}
\end{equation}%
while, furthermore, the \emph{entropic equality condition}%
\begin{equation}
\frac{\partial }{\partial t}S(\rho _{1}(t))=0\Leftrightarrow \rho _{1}(%
\mathbf{x}_{1},t)=\rho _{1M}^{(N)}(\mathbf{v}_{1})  \label{entropic equality}
\end{equation}%
should hold. The latter equation implies therefore that, provided $\rho
_{1}(t)$ and $S(\rho _{1}(t))$ exist globally \cite{Villani}, then
necessarily $lim_{t\rightarrow +\infty }\rho _{1}(\mathbf{x}_{1},t)=\rho
_{1M}(\mathbf{v}_{1}),$ with $\rho _{1M}(\mathbf{v}_{1})$ denoting the
stationary and spatially-uniform Maxwellian PDF (\ref{MAXWELLIAN-PDF}).

In this reference, however, the question arises of the precise
characterization of the concept of irreversibility, i.e., whether it should
be regarded as a purely macroscopic phenomenon ("macroscopic
irreversibility"), i.e., affecting only the BS entropy $S(\rho _{1}(t))$
through the Boltzmann H-theorem indicated above,\ or microscopic in the
sense that the same Boltzmann equation should be considered as irreversible
("microscopic irreversibility"). Thus, in principle, in the second case the
further issue emerges of the possible physical origin of microscopic
irreversibility in special reference to the Lanford's derivation of the
Boltzmann equation and subsequent related comments discussed respectively by
Uffink and Valente and Ardourel in Refs. \cite{Uffink2015} and \cite%
{Ardourel} (see also Drory \cite{Droty2008}).

However, as shown in Ref.\cite{noi11}, the Boltzmann equation is actually $%
\mathcal{TR}-$symmetric. Such a conclusion is of basic importance since it
overcomes the so-called Loschmidt paradox,\ i.e., the objection raised by
Loschmidt in 1876 \cite{Loschmidt1876} regarding the original Boltzmann
formulation of his namesake kinetic equation and H-theorem \cite%
{Boltzmann1972}. In fact, Loschmidt claimed that the Boltzmann H-theorem
inequality should change sign under time reversal and thus violate the
microscopic time-reversibility of the underlying hard-sphere classical
dynamical system. In his long-pondered reply given in 1896 \cite%
{Boltzmann1896} Boltzmann himself introduced what was later referred to as
the\textbf{\ }modified form of the Boltzmann H-theorem \cite{Ehrenfest}.

The key implication is therefore that, in contrast to Boltzmann's own
statement and the traditional subsequent mainstream literature
interpretation (see for example by Cercignani, Lebowitz in Refs. \cite%
{Cercignani1982,Lebowitz1993} and more recently the review given by
Gallavotti \cite{Gallavotti2014}), the Boltzmann H-theorem indicated -
together with the modified form indicated above - cannot be interpreted as
an intrinsic irreversibility property occurring at the microscopic level,
namely holding for the Boltzmann equation itself. On the contrary,
consistent with the physical interpretation of the Loschmidt paradox
provided in Ref.\cite{noi11}, this must be regarded only as\textbf{\ }%
\textit{property of macroscopic irreversibility}\textbf{\ }(or PMI) of the $%
1-$body PDF solution of the Boltzmann equation. In other words, the
Boltzmann inequality (\ref{Entropic enequality}) necessarily holds\emph{\
independent of the orientation of the time axis}\textbf{\ }(arrow of time)
and therefore cannot represent a true (i.e., microscopic) property which as
such should uniquely determine the arrow of time.

Nevertheless, the possible realization of either PMI or DKE is more subtle.
In fact they actually depend in a critical way on the prescription of the
functional class $\left\{ \rho ^{(N)}(\mathbf{x}_{1},t)\right\} ,$ so that
their occurrence is actually non-mandatory. Indeed, both cannot occur - in
principle also for Boltzmann and Grad kinetic theories - if the $N-$body
probability density function $\rho ^{(N)}(\mathbf{x},t)$ is identified with
the deterministic $N-$body PDF \cite{noi1}, namely the $N-$body phase-space
Dirac delta. This is defined as $\delta (\mathbf{x}-\mathbf{x}(t))\equiv
\prod\limits_{1=1,N}\delta (\mathbf{x}_{i}-\mathbf{x}_{i}(t)),$ with $%
\mathbf{x}\equiv \left\{ \mathbf{x}_{1},...,\mathbf{x}_{N}\right\} $
denoting the state of the $N-$body system and $\mathbf{x}(t)\equiv \left\{
\mathbf{x}_{1}(t),...,\mathbf{x}_{N}(t)\right\} $ is the image of an
arbitrary initial state $\mathbf{x}(t_{o})\equiv \mathbf{x}_{o}$ generated
by the same $N-$body CDS. That such a PDF necessarily must realize an
admissible particular solution of the $N-$body Liouville equation follows,
in fact, as a straightforward consequence of the axioms of classical
statistical mechanics \cite{noi1}.

Despite these premises, however, the case of a finite Boltzmann-Sinai CDS,
which is characterized by a finite number of particles $N$\ and/or a
finite-size of the hard spheres and/or a dense or locally-dense system, is
more subtle and - as explained below - even unprecedented since it has
actually remained unsolved to date. The reasons are as follows. First,
Boltzmann and Grad kinetic theories are inapplicable to the finite
Boltzmann-Sinai CDS. Second, the Boltzmann-Shannon entropy associated with
an arbitrary particular solution $\rho ^{(N)}(t)\equiv \rho ^{(N)}(\mathbf{x}%
_{1},t)$ of the Master kinetic equation, i.e., the functional $S(\rho
_{1}^{(N)}(t))\equiv M_{X_{E}}(\rho _{1}^{(N)}(t)),$ in contrast to $S(\rho
_{1}(t))\equiv M_{X_{E}}(\rho _{1}(t)),$ is exactly conserved in the sense
that identically%
\begin{equation}
\frac{\partial S(\rho _{1}^{(N)}(t))}{\partial t}\equiv 0
\label{constant-H theorem}
\end{equation}%
must hold. As a consequence the validity itself of Boltzmann H-theorem
breaks down in the case of the Master kinetic equation. Third, an additional
motivation is provided by the conjecture that both PMI and DKE might occur
only if the Boltzmann-Grad limit is actually performed, i.e., only in
validity of Boltzmann equation and H-theorem.

Hence the question which arises is whether in the case of a finite
Boltzmann-Sinai CDS the phenomenon of DKE may still arise. Strong
indications seem to be hinting at such a possibility. In this regard the
example-case which refers to the statistical description of a Navier-Stokes
fluid described by the incompressible Navier-Stokes equations (INSE) in
terms of the Master kinetic equation is relevant and suggests that this may
be indeed the case. In fact, thanks also to comparisons with the mean-field
inverse kinetic approach to INSE \cite{ikt1}, in such a case the decay of
the fluid velocity field occurring in a bounded domain necessarily demands
the existence of DKE. In other words, in the limit $t\rightarrow +\infty $
the $1-$body PDF must decay uniformly to the stationary and
spatially-uniform Maxwellian PDF (\ref{MAXWELLIAN-PDF}).

However, besides the construction of the kinetic equation appropriate for
such a case, a further unsolved issue lies in the determination of the
functional class $\left\{ \rho _{1}^{(N)}(\mathbf{x}_{1},t)\right\} $ for
which both PMI and DKE should/might be realized. In particular, the possible
occurrence of both PMI and DKE should correspond to suitably-smooth, but
nonetheless still arbitrary, initial conditions $\left\{ \rho _{1}^{(N)}(%
\mathbf{x}_{1},t_{o})\right\} $. These should warrant that in the limit $%
t\rightarrow +\infty ,$ $\rho _{1}^{(N)}(\mathbf{x}_{1},t)$ uniformly
converges to the spatially-homogeneous and stationary Maxwellian PDF $\rho
_{1M}(\mathbf{v}_{1})$ (\ref{MAXWELLIAN-PDF}). Such a result, however, is
highly non-trivial since it should rely on the establishment of a global
existence theorem for the same $1-$body PDF $\rho _{1}^{(N)}(\mathbf{x}%
_{1},t)$ - namely holding in the whole time axis $I\equiv
\mathbb{R}
$, besides the same $1-$body phase space $\Gamma _{1}$ - for the involved
kinetic equation which is associated with the $S_{N}-$CDS. In the context of
the Boltzmann equation in particular, despite almost-endless efforts this
task has actually not been accomplished yet, the obstacle being
intrinsically related to the asymptotic nature of the Boltzmann equation
\cite{noi7}. In fact for the same equation it is not known\ in satisfactory
generality whether smooth enough solutions of the same equation exist which
satisfy the $H-$theorem inequality and decay asymptotically to kinetic
equilibrium \cite{Cercignani1982,Villani}.

\subsection{1B - Goals and organization of the paper}

Based on these premises, the crucial new results that we intend to display
in this paper concern the \emph{proof-of-principle}\textbf{\ }of two
phenomena which are expected to characterize the statistical description of
finite $N-$body hard-sphere systems and therefore should lay at the very
foundation of classical statistical mechanics and kinetic theory alike.%
\textbf{\ }These are related to the physical conditions for the possible
occurrence of both PMI and the consequent one represented by the possible
occurrence of DKE which should characterize the kinetic PDF in these
systems. These phenomena are well known to occur in the case of dilute
hard-sphere systems, i.e., in the Boltzmann-Grad limit. In particular, for
an exhaustive treatment of the related issues which arise in the context of
the ab initio theory we refer to discussions reported in Ref. \cite{noi11}.
Nevertheless, as indicated above, their existence in the case of finite
hard-sphere systems is partly motivated by a previous investigation dealing
with the kinetic description incompressible Navier-Stokes granular fluids
\cite{noi8}.

Therefore, main goal of the paper is\ to show that these properties actually
emerge as necessary implications of the ab initio theory of CSM.
Incidentally, in doing so, the Master kinetic equation must be necessarily
adopted. In fact, the finiteness requirement on the $S_{N}-$CDS rules out
for further possible consideration either the Boltzmann or the Enskog kinetic%
\textbf{\ }equations, these equations being inapplicable to the treatment of
systems of this type \cite{noi3}. Specifically, in the following the case $%
N>2$\ is considered everywhere, which is by far the most physically-relevant
one. In this occurrence, in fact, non-trivial $2-$body occupation
coefficients arise (see related notations which are applicable for $N>2$\
recalled in Appendices A and B below).\textbf{\ }For completeness the case $%
N=2$\ is nevertheless briefly discussed in Appendix D.

For this purpose, first, in Section 2, the MKI functional is explicitly
determined. We display in particular its construction method (see \emph{%
No.\#1- \#4 MKI Prescriptions}). Based on the theory of the Master kinetic
equation earlier developed \cite{noi3} and suitable integral and
differential identities (see Appendices A, B and C), the properties of the
MKI functional are investigated. These concern in particular the
establishment of appropriate inequalities holding for the same functional
(THM.1, subsection 2A), the signature of the time derivative of the same
functional (THM.2, subsection 2B) and the property of DKE holding for a
suitable class of $1-$body PDFs (THM.3, subsection 2C). In the subsequent
Sections 3 and 4, the issue of the consistency of the phenomena of PMI and
DKE with microscopic dynamics is posed together with the physical
interpretation and implications of the theory. The\ goal is to investigate
the relationship of the DKE-theory developed here with the microscopic
reversibility principle and\ the Poincar\'{e} recurrence theorem. Finally in
Section 5 the conclusions of the paper are drawn and possible
applications/developments of the theory are pointed out.

\section{2 - Axiomatic prescription of the MKI functional}

In view of the considerations given above in this section the problem is
posed of the explicit realization of the MKI functional in terms of suitable
axiomatic prescriptions. The same functional, denoted\textbf{\ }$I_{M}(\rho
_{1}^{(N)}(t)),$\textbf{\ }should depend on the $1-$body PDF $\rho
_{1}^{(N)}(t),$\ with $\rho _{1}^{(N)}(t)\equiv \rho _{1}^{(N)}(\mathbf{x}%
_{1},t)$ being identified with a particular solution of the Master kinetic
equation (see Eq.(\ref{App-1}) in Appendix A holding for $N>2$\ and Appendix
D for the case $N=2$).

Unlike Boltzmann kinetic equation, the Master kinetic equation actually
deals with the treatment of finite hard-sphere $N-$body systems,\ i.e., in
which both the number of particles $N$\ and their diameter $\sigma $\ remain
finite \cite{noi3}.\ To achieve such a goal suitably-prescribed physical
collision boundary conditions (CBC) of the $N-$body PDF need to be adopted.\
More precisely, this concerns the prescription\textbf{\ }for arbitrary
collision events of the relationship between incoming ($-$) and outgoing ($+$%
) PDFs, i.e., respectively the left and right limits $\rho ^{(\pm )(N)}(%
\mathbf{x}^{(\pm )}(t_{i}),t_{i})=\lim_{t\rightarrow t_{i}^{(\pm )}}\rho
^{(N)}(\mathbf{x}(t),t),$ with $\mathbf{x}^{(\pm
)}(t_{i})=\lim_{t\rightarrow t_{i}^{(\pm )}}\mathbf{x}(t)$ denoting the
corresponding incoming ($-$) and outgoing ($+$) states. In particular, upon
invoking due to causality the assumption of left-continuity,\ i.e., the
requirement%
\begin{equation}
\rho ^{(-)(N)}(\mathbf{x}^{(-)}(t_{i}),t_{i})\equiv \rho ^{(N)}(\mathbf{x}%
^{(-)}(t_{i}),t_{i}),  \label{LEFT-CONTINUITY-1}
\end{equation}%
the incoming PDF is required to coincide with the same $N-$body PDF
evaluated in terms of the incoming state and time \cite{noi1,noi3}. Hence,
as recalled in Appendix C (see also Ref. \cite{noi2}) from Eq.(\ref{bbb1})
if follows that the so-called causal form of the modified collision boundary
condition (MCBC \cite{noi2})%
\begin{equation}
\rho ^{(+)(N)}(\mathbf{x}^{(+)}(t_{i}),t_{i})=\rho ^{(N)}(\mathbf{x}%
^{(+)}(t_{i}),t_{i})  \label{CBC-2}
\end{equation}%
is mandatory. A further important requirement concerns precisely setting\
also the related\textbf{\ }\emph{functional class of admissible solutions} $%
\left\{ \rho _{1}^{(N)}(\mathbf{x}_{1},t)\right\} $\ in such a\textbf{\ }way
that, besides $\rho _{1}^{(N)}(t)$, also\ the same functional $I_{M}(\rho
_{1}^{(N)}(t))$\ exists globally for arbitrary $t\in I\equiv
\mathbb{R}
$. For definiteness, we shall consider for this purpose the case of $1-$body
PDFs which satisfy the initial condition%
\begin{equation}
\rho _{1}^{(N)}(t_{o})\equiv \rho _{1}^{(N)}(\mathbf{x}_{1},t_{o})=\rho
_{1(o)}^{(N)}(\mathbf{x}_{1}),  \label{initial condition}
\end{equation}%
with $\rho _{1(o)}^{(N)}(\mathbf{x}_{1})$\textbf{\ }belonging to the
functional class of \emph{stochastic }$1-$\emph{body PDFs }$\left\{ \rho
_{1}^{(N)}(t_{o})\right\} $. For a generic $t\in I$ belonging to the time
axis $I\equiv
\mathbb{R}
$ this is the ensemble of $1-$body PDFs $\rho _{1}^{(N)}(t)$ which are
respectively: A) smoothly differentiable; B) strictly positive; C) summable,
in the sense that the velocity - or phase-space - moments for the same PDF
exist which correspond either to arbitrary monomial functions of $\mathbf{v}%
_{1}$\ (or its components $v_{1i},$\ for $i=1,2,3$) or to the entropy
density $\ln \rho _{1}^{(N)}(t),$\ thus yielding the Boltzmann-Shannon (BS)
entropy evaluated in terms of $\rho _{1}^{(N)}(t)$.

Concerning the choice of the setting $\left\{ \rho _{1}^{(N)}(t_{o})\right\}
$ the following remarks are in order. As a first remark, the previous
requirements A), B) and C) for $\left\{ \rho _{1}^{(N)}(t_{o})\right\} ,$
together with validity of MCBC (\ref{CBC-2}), actually\ should warrant that
the corresponding solution of the Master kinetic equation $\rho
_{1}^{(N)}(t) $\ exists globally in the extended phase-space\textbf{\ }$%
\left( \mathbf{x}_{1},t\right) \in \Gamma _{1}\times I$ and that for all $%
t\in I$\textbf{\ }the same PDF belongs to the class of stochastic PDFs $%
\left\{ \rho _{1}^{(N)}(\mathbf{x}_{1},t)\right\} $ indicated above and also
fulfills identically the constant H-theorem\textbf{\ }(\ref{constant-H
theorem}). Indeed, one can show \cite{noi7} that global existence of
solutions for the Master kinetic equation follows in elementary way from the
$N-$body Liouville equation. Indeed, an arbitrary $1-$body PDF which is a
particular solution of the Master kinetic equation realizes by construction
also a particular factorized solution of the $N-$body Liouville equation,
i.e., of the $N-$body PDF \cite{noi3}. The same PDF evolves uniquely in time
along arbitrary phase-space Lagrangian trajectories, its Lagrangian time
evolution being determined at arbitrary collision times by MCBC (\ref{CBC-2}%
) \cite{noi7}.

\emph{As a second remark, }the validity of assumptions A) and B) for\textbf{%
\ }$\rho _{1(o)}^{(N)}(\mathbf{x}_{1})$ and $\rho _{1}^{(N)}(\mathbf{x}%
_{1},t)$ implies also suitable assumptions to apply all $t\in I$\ \ for the
local characteristic scale-length $L_{\rho }$\ which characterize the same
PDF $\rho _{1}^{(N)}(\mathbf{x}_{1},t).$ More precisely, this is associated
with the spatial variations of the $1-$body PDF prescribed as
\begin{equation}
L_{\rho }(t)=\inf_{\mathbf{x}_{1}\in \Gamma _{1}}\left\{ \left\vert \frac{%
\partial \ln \rho _{1}^{(N)}(\mathbf{x}_{1},t)}{\partial \mathbf{r}_{1}}%
\right\vert ^{-1}\right\} ,  \label{b-lunghezza}
\end{equation}%
which necessarily assumed non-zero at all time $t\in I$. Hence $\frac{%
\partial \ln \rho _{1}^{(N)}\left( \mathbf{x}_{1},t\right) }{\partial
\mathbf{r}_{1}}$ is assumed to be bounded for all $\left( \mathbf{x}%
_{1},t\right) $ spanning the extended $1-$body phase space $\Gamma
_{1(1)}\times I$ .

Finally, as a third remark (see also the further related discussion in THM.2
below), the previous requirements are expected to warrant also the global
existence of the MKI functional $I_{M}(\rho _{1}^{(N)}(t))$,\ so that $%
\left\{ \rho _{1}^{(N)}(\mathbf{x}_{1},t)\right\} $ effectively realizes the
functional class of admissible solutions indicated above.

Given these premises let us pose now the problem of the identification of
the functional $I_{M}\left( \rho _{1o}^{(N)}(\mathbf{x}_{1})\right) $, based
on the introduction of 'ad hoc' physical requirements, to be referred to
here as \emph{MKI Prescriptions No.\#1-\#4}.\ The prescriptions are as
follows:

\begin{itemize}
\item \emph{MKI Prescription No.\#1:\ }the first one is that the functional $%
I_{M}(\rho _{1}^{(N)}(t))$ should be determined in such a way that the
existence of $I_{M}\left( \rho _{1o}^{(N)}(\mathbf{x}_{1})\right) $ at a
suitable initial time $t_{o}\in I$ should warrant also that $I_{M}(\rho
_{1}^{(N)}(t))$ must necessarily exist globally in the future, i.e., for all
$t\geq t_{o}.$ As a consequence the functional class $\left\{ \rho
_{1}^{(N)}(\mathbf{x}_{1},t)\right\} $\ must be suitably prescribed.

\item \emph{MKI Prescription No.\#2: }second, we shall require that $%
I_{M}(\rho _{1}^{(N)}(t))$ is real, non-negative and bounded in the sense
that%
\begin{equation}
0\leq I_{M}(\rho _{1}^{(N)}(t))\leq 1.  \label{MKI-1}
\end{equation}%
This implies that $I_{M}(\rho _{1}^{(N)}(t))$ can be interpreted as an
information measure associated with the $1-$body PDF $\rho
_{1}^{(N)}(t)\equiv \rho _{1}^{(N)}(\mathbf{x}_{1},t)$. For this reason the
previous inequalities will be referred to as\ \emph{information-measure
inequalities}.

\item \emph{MKI Prescription No.\#3:\ }third, for consistency with the
property of macroscopic irreversibility, $I_{M}(\rho _{1}^{(N)}(t))$ is
prescribed in terms of a smoothly time-differentiable and monotonically
time-decreasing functional in the sense that in the same time-subset the
inequality:%
\begin{equation}
\frac{\partial }{\partial t}I_{M}(\rho _{1}^{(N)}(t))\leq 0  \label{MKI-2}
\end{equation}%
should identically apply $\forall t\geq t_{o},$ so that%
\begin{equation}
0\leq I_{M}(\rho _{1}^{(N)}(t))\leq I_{M}\left( \rho _{1o}^{(N)}(\mathbf{x}%
_{1})\right) \leq 1.  \label{MKI-2bis}
\end{equation}%
This implies that $I_{M}(\rho _{1}^{(N)}(t))$ is also globally defined for
all $t\in I\equiv
\mathbb{R}
$ with $t\gtrsim t_{o}.$ In addition, if $\left. \frac{\partial }{\partial t}%
I_{M}(\rho _{1}^{(N)}(t))\right\vert _{t=t_{o}}\neq 0$, without loss of
generality its initial value $I_{M}\left( \rho _{1o}^{(N)}(\mathbf{x}%
_{1})\right) $ can always be set such that%
\begin{equation}
I_{M}\left( \rho _{1o}^{(N)}(\mathbf{x}_{1})\right) =1.  \label{MKI-2ter}
\end{equation}

\item \emph{MKI Prescription No.\#4:\ }fourth, in order to warrant the
existence of DKE we shall require the functional $I_{M}(\rho _{1}^{(N)}(t))$
to be prescribed in such a way that at an arbitrary time $t\in I,$ with $%
t\gtrsim t_{o},$ the vanishing of both $I_{M}(\rho _{1}^{(N)}(t))$ and its
time derivative $\frac{\partial }{\partial t}I_{M}(\rho _{1}^{(N)}(t))$
should occur if and only if the $1-$body PDF solution of the Master kinetic
equation coincides with kinetic equilibrium. As a consequence, for the
functional $I_{M}(\rho _{1}^{(N)}(t))$ the following propositions should be
equivalent%
\begin{equation}
\left\{
\begin{array}{c}
I_{M}(\rho _{1}^{(N)}(t))=0 \\
\frac{\partial }{\partial t}I_{M}(\rho _{1}^{(N)}(t))=0%
\end{array}%
\Leftrightarrow \rho _{1}^{(N)}(\mathbf{x}_{1},t)\equiv \rho _{1M}^{(N)}(%
\mathbf{v}_{1}),\right.  \label{MKI-4a}
\end{equation}%
with $\rho _{1M}^{(N)}(\mathbf{v}_{1})$ being a kinetic equilibrium PDF of
the form (\ref{MAXWELLIAN-PDF}).
\end{itemize}

The immediate obvious implication of the previous prescriptions is that
\textbf{- }provided a non-trivial realization of the MKI can be found in the
functional class $\left\{ \rho _{1o}^{(N)}(\mathbf{x}_{1})\right\} $\textbf{%
\ - }the existence of both PMI and DKE for the Master kinetic equation%
\textbf{\ }is actually established. In the sequel the goal is to show, in
particular, that the MKI functional can be identified by means of the
prescription%
\begin{equation}
I_{M}(\rho _{1}^{(N)}(t),\mathbf{b})\equiv \frac{K_{M}(\rho _{1}^{(N)}(t),%
\mathbf{b})}{K_{Mo}},  \label{MKI-functional-1}
\end{equation}%
where $K_{M}(\rho _{1}^{(N)}(t),\mathbf{b})$ and $K_{Mo}$ denote
respectively a suitable (and possibly non-unique) moment-dependent
phase-space functional and an appropriate normalization constant to be
chosen in such a way to satisfy all the MKI prescriptions indicated above.%
\textbf{\ }In particular, as shown below, an admissible choice for $%
K_{M}(\rho _{1}^{(N)}(t),b)$ and $K_{Mo}$\ is provided by%
\begin{equation}
\left\{
\begin{array}{c}
K_{M}(\rho _{1}^{(N)}(t),\mathbf{b})=-\int\limits_{\Gamma _{1(1)}}d\mathbf{x}%
_{1}\overline{\Theta }_{1}^{(\partial \Omega )}(\overline{\mathbf{r}}_{1})M(%
\mathbf{v}_{1},\mathbf{b})\frac{\rho _{1}^{(N)}(\mathbf{x}_{1},t)}{\widehat{%
\rho }_{1}^{(N)}(\mathbf{x}_{1},t)}\frac{\partial ^{2}\widehat{\rho }%
_{1}^{(N)}(\mathbf{x}_{1},t)}{\partial \mathbf{r}_{1}\cdot \partial \mathbf{r%
}_{1}}, \\
K_{Mo}=\sup \left\{ 1,K_{M}(\rho _{1o}^{(N)}(\mathbf{x}_{1}),\mathbf{b}%
)\right\} ,%
\end{array}%
\right.  \label{PRESCRIPTIONS FOR K_M and K_Mo}
\end{equation}%
while $M(\mathbf{v}_{1},\mathbf{b})$ denotes the \emph{directional kinetic
energy} (along the unit vector $\mathbf{b}$) carried by particle $1$, namely
the dynamical variable%
\begin{equation}
M(\mathbf{v}_{1},\mathbf{b})\equiv \left( \mathbf{v}_{1}\cdot \mathbf{b}%
\right) ^{2},  \label{MOMENT-1}
\end{equation}%
with $\mathbf{b}$ denoting a still arbitrary constant unit vector. Hence,
\begin{equation}
M(\mathbf{v}_{1},\mathbf{v}_{2},\mathbf{b})=\frac{1}{2}\left[ M(\mathbf{v}%
_{1},\mathbf{b})+M(\mathbf{v}_{2},\mathbf{b})\right]
\label{TOTAL DIRECTIONAL ENERGY}
\end{equation}%
identifies the corresponding \emph{total directional kinetic energy} carried
by particles $1$ and $2$. Here the remaining notation is standard. Thus, $%
\rho _{1}^{(N)}(t)\equiv \rho _{1}^{(N)}(\mathbf{x}_{1},t),$ $\rho
_{1o}^{(N)}(\mathbf{x}_{1})$ and $\widehat{\rho }_{1}^{(N)}(t)\equiv
\widehat{\rho }_{1}^{(N)}(\mathbf{x}_{1},t)$ are respectively the $1-$body
PDF solution of the initial problem associated with the Master kinetic
equation (see Eq.(\ref{App-1}) in Appendix A), the initial PDF and the
renormalized $1-$body PDF
\begin{equation}
\widehat{\rho }_{1}^{(N)}(\mathbf{x}_{1},t)\equiv \frac{\rho _{1}^{(N)}(%
\mathbf{x}_{1},t)}{k_{1}^{(N)}(\mathbf{r}_{1},t)},  \label{App-00}
\end{equation}%
while furthermore $k_{1}^{(N)}(\mathbf{r}_{1},t)$ is the $1-$body occupation
coefficient recalled in Appendix B (see Eq.(\ref{App-4})). As a consequence
in the previous equation it follows that $\frac{\rho _{1}^{(N)}(\mathbf{x}%
_{1},t)}{\widehat{\rho }_{1}^{(N)}(\mathbf{x}_{1},t)}\equiv k_{1}^{(N)}(%
\mathbf{r}_{1},t)$. Furthermore, $\overline{\Theta }_{1}^{(\partial \Omega
)}(\overline{\mathbf{r}}_{1})$ is the boundary theta-function given by Eq.(%
\ref{boundary theta function}) (see Appendix A). Finally, regarding the
initial value\textbf{\ }$K_{M}(\rho _{1o}^{(N)}(\mathbf{x}_{1}),\mathbf{b})$
it follows that if respectively\textbf{\ }$K_{M}(\rho _{1o}^{(N)}(\mathbf{x}%
_{1}),\mathbf{b})\geq 1$ or
\begin{equation}
0\leq K_{M}(\rho _{1o}^{(N)}(\mathbf{x}_{1}),\mathbf{b})<1,
\label{double inequality}
\end{equation}%
then correspondingly one obtains, consistent with (\ref{MKI-2bis}), that the
initial value of MKI functional $I_{M}(\rho _{1o}^{(N)},\mathbf{b})$ is%
\textbf{\ }%
\begin{equation}
I_{M}(\rho _{1o}^{(N)},\mathbf{b})=\left\{
\begin{array}{c}
1, \\
K_{M}(\rho _{1o}^{(N)}(\mathbf{x}_{1}),\mathbf{b}).%
\end{array}%
\right.  \label{MKI-6}
\end{equation}

\subsection{2A - Proof of the non-negativity of the MKI information measure}

The strategy adopted for the proof of the MKI Prescriptions No.\#1 and
No.\#2 is to show initially the validity of the information-measure left
inequality in Eq.(\ref{MKI-1}), namely that $I_{M}(\rho _{1}^{(N)}(t),%
\mathbf{b})$ cannot acquire negative values for arbitrary $t\geq t_{o}$. The
result is established by the following theorem.

\bigskip

\textbf{THM. 1 - Non-negativity of }$K_{M}(\rho _{1o}^{(N)}(\mathbf{x}_{1}),%
\mathbf{b}),$ $K_{M}(\rho _{1}^{(N)}(t),\mathbf{b})$ \textbf{and} $%
I_{M}(\rho _{1}^{(N)}(t),\mathbf{b})$

\emph{Let us assume that }$\rho _{1}^{(N)}(\mathbf{x}_{1},t)$\emph{\ is an
arbitrary stochastic and suitably smoothly-differentiable, particular
solution of the Master kinetic equation (\ref{App-1}) prescribed so that\
the integral (\ref{PRESCRIPTIONS FOR K_M and K_Mo})\ expressed in terms of
the initial PDF, namely}\textbf{\ }$K_{M}(\rho _{1o}^{(N)}(\mathbf{x}_{1}),%
\mathbf{b}),$ \emph{is non-vanishing. Then, it follows necessarily that:}

\begin{itemize}
\item \emph{Proposition P1}$_{1}:$%
\begin{equation}
K_{M}(\rho _{1o}^{(N)}(\mathbf{x}_{1}),\mathbf{b})>0.  \label{P3-1-1}
\end{equation}

\item \emph{Proposition P1}$_{2}:$ \emph{the corresponding time-evolved
functional} $K_{M}(\rho _{1}^{(N)}(t),\mathbf{b})$ \emph{for all }$t\in I$%
\emph{\ with }$t>t_{o}$ \emph{is such that}
\begin{equation}
K_{M}(\rho _{1}^{(N)}(t),\mathbf{b})\geq 0.  \label{P3-1-2}
\end{equation}

\item \emph{Proposition P1}$_{3}:$ \emph{for all }$t\in I$\emph{\ with }$%
t>t_{o}$\emph{\ the functional }$I_{M}(\rho _{1}^{(N)}(t),\mathbf{b})$ \emph{%
fulfills the inequality}%
\begin{equation}
I_{M}(\rho _{1}^{(N)}(t),\mathbf{b})\geq 0.  \label{P3-1-3}
\end{equation}

\item \emph{Proposition P1}$_{4}:$ \emph{the following necessary and
sufficient condition holds at a given time }$t\in I$ \emph{with} $t\geq
t_{o}:$%
\begin{equation}
K_{M}(\rho _{1}^{(N)}(t),\mathbf{b})=0\Leftrightarrow \rho _{1}^{(N)}(%
\mathbf{x}_{1},t)\equiv \rho _{1M}^{(N)}(\mathbf{v}_{1}).  \label{P3-1-4}
\end{equation}
\end{itemize}

\emph{Proof - }One first notices that $K_{M}(\rho _{1}^{(N)}(t),\mathbf{b})$
can be equivalently written in the form%
\begin{equation}
K_{M}(\rho _{1}^{(N)}(t),\mathbf{b})\equiv -\int\limits_{\Gamma _{1(1)}}d%
\mathbf{x}_{1}\overline{\Theta }_{1}^{(\partial \Omega )}(\overline{\mathbf{r%
}}_{1})M(\mathbf{v}_{1},\mathbf{b})k_{1}^{(N)}(\mathbf{r}_{1},t)\frac{%
\partial ^{2}\widehat{\rho }_{1}^{(N)}(\mathbf{x}_{1},t)}{\partial \mathbf{r}%
_{1}\cdot \partial \mathbf{r}_{1}},
\end{equation}%
where in order that the same functional exists it is obvious that the
renormalized $1-$body PDF $\widehat{\rho }_{1}^{(N)}(\mathbf{x}_{1},t)$\
must be of class $C^{(2)}$. Integrating by parts and noting that the
gradient term $\frac{\partial \overline{\Theta }_{1}^{(\partial \Omega )}(%
\overline{\mathbf{r}})}{\partial \mathbf{r}_{1}}$ gives a vanishing
contribution to the phase-space integral, this yields equivalently%
\begin{equation}
K_{M}(\rho _{1}^{(N)}(t),\mathbf{b})\equiv \int\limits_{\Gamma _{1(1)}}d%
\mathbf{x}_{1}\overline{\Theta }_{1}^{(\partial \Omega )}(\overline{\mathbf{r%
}}_{1})M(\mathbf{v}_{1},\mathbf{b})\frac{\partial k_{1}^{(N)}(\mathbf{r}%
_{1},t)}{\partial \mathbf{r}_{1}}\cdot \frac{\partial \widehat{\rho }%
_{1}^{(N)}(\mathbf{x}_{1},t)}{\partial \mathbf{r}_{1}}.
\end{equation}%
Therefore, upon invoking Eq.(\ref{App-X1}) reported in Appendix B, direct
substitution delivers%
\begin{eqnarray}
&&\left. K_{M}(\rho _{1}^{(N)}(t),\mathbf{b})=\left( N-1\right)
\int\limits_{\Gamma _{1(1)}}d\mathbf{x}_{1}\overline{\Theta }_{1}^{(\partial
\Omega )}(\overline{\mathbf{r}}_{1})M(\mathbf{v}_{1},\mathbf{b})\frac{%
\partial \widehat{\rho }_{1}^{(N)}(\mathbf{x}_{1},t)}{\partial \mathbf{r}_{1}%
}\cdot \int\limits_{\Gamma _{1(2)}}d\mathbf{x}_{2}\mathbf{n}_{12}\times
\right.  \notag \\
&&\left. \delta \left( \left\vert \mathbf{r}_{2}-\mathbf{r}_{1}\right\vert
-\sigma \right) \overline{\Theta }_{2}^{(\partial \Omega )}(\overline{%
\mathbf{r}}_{2})\widehat{\rho }_{1}^{(N)}(\mathbf{x}_{2},t)k_{2}^{(N)}(%
\mathbf{r}_{1},\mathbf{r}_{2},t).\right.  \label{RAPPRESENTAZIONE}
\end{eqnarray}%
Next, invoking the identity $\mathbf{n}_{12}\delta \left( \left\vert \mathbf{%
r}_{2}-\mathbf{r}_{1}\right\vert -\sigma \right) =-\frac{\partial }{\partial
\mathbf{r}_{2}}\overline{\Theta }\left( \left\vert \mathbf{r}_{2}-\mathbf{r}%
_{1}\right\vert -\sigma \right) $ and noting again that $\frac{\partial }{%
\partial \mathbf{r}_{2}}\overline{\Theta }_{2}^{(\partial \Omega )}(%
\overline{\mathbf{r}})$ gives vanishing contribution,\textbf{\ }one can
perform a further integration by parts with respect to $\mathbf{r}_{2}$.
This permits to cast the rhs of previous equation in the form%
\begin{equation}
K_{M}(\rho _{1}^{(N)}(t),\mathbf{b})\equiv K_{M}^{(1)}(\rho _{1}^{(N)}(%
\mathbf{x}_{1},t),\mathbf{b})+\Delta K_{M}^{(1)}(\rho _{1}^{(N)}(\mathbf{x}%
_{1},t),\mathbf{b}).  \label{LAST}
\end{equation}%
Here the two terms on the rhs of Eq.(\ref{LAST}) are defined as follows: 1)
the first term $K_{M}^{(1)}(\rho _{1}^{(N)}(t),\mathbf{b})$ is symmetric and
non-negative, so that it can be expressed so to carry the total directional
kinetic energy $M(\mathbf{v}_{1},\mathbf{v}_{2},\mathbf{b})$ of particles $1$
and $2$ (see Eq.(\ref{TOTAL DIRECTIONAL ENERGY})). Hence, it takes the form
\begin{equation}
\begin{array}{c}
K_{M}^{(1)}(\rho _{1}^{(N)}(\mathbf{x}_{1},t),\mathbf{b})=\left( N-1\right)
\int\limits_{\Gamma _{1(1)}}d\mathbf{x}_{1}\int\limits_{\Gamma _{1(2)}}d%
\mathbf{x}_{2}\overline{\Theta }_{1}^{(\partial \Omega )}(\overline{\mathbf{r%
}}_{1})\overline{\Theta }_{2}^{(\partial \Omega )}(\overline{\mathbf{r}}%
_{2})\times \\
\frac{\partial \widehat{\rho }_{1}^{(N)}(\mathbf{x}_{1},t)}{\partial \mathbf{%
r}_{1}}\cdot \frac{\partial \widehat{\rho }_{1}^{(N)}(\mathbf{x}_{2},t)}{%
\partial \mathbf{r}_{2}}k_{2}^{(N)}(\mathbf{r}_{1},\mathbf{r}_{2},t)M(%
\mathbf{v}_{1},\mathbf{v}_{2},\mathbf{b})\overline{\Theta }\left( \left\vert
\mathbf{r}_{2}-\mathbf{r}_{1}\right\vert -\sigma \right) .%
\end{array}
\label{K^1}
\end{equation}%
2) The second term\textbf{\ }$\Delta K_{M}^{(1)}(\rho _{1}^{(N)}(\mathbf{x}%
_{1},t),\mathbf{b})$ reads instead
\begin{eqnarray}
&&\Delta K_{M}^{(1)}(\rho _{1}^{(N)}(\mathbf{x}_{1},t),\mathbf{b})\equiv
\left( N-1\right) \int\limits_{\Gamma _{1(1)}}d\mathbf{x}_{1}\overline{%
\Theta }_{1}^{(\partial \Omega )}(\overline{\mathbf{r}}_{1})\frac{\partial
\widehat{\rho }_{1}^{(N)}(\mathbf{x}_{1},t)}{\partial \mathbf{r}_{1}}\cdot
\int\limits_{\Gamma _{1(2)}}d\mathbf{x}_{2}\times  \notag \\
&&\overline{\Theta }_{2}^{(\partial \Omega )}(\overline{\mathbf{r}}_{2})M(%
\mathbf{v}_{1},\mathbf{b})\overline{\Theta }\left( \left\vert \mathbf{r}_{2}-%
\mathbf{r}_{1}\right\vert -\sigma \right) \widehat{\rho }_{1}^{(N)}(\mathbf{x%
}_{2},t)\frac{\partial }{\partial \mathbf{r}_{2}}k_{2}^{(N)}(\mathbf{r}_{1},%
\mathbf{r}_{2},t),  \label{AK^1}
\end{eqnarray}%
where $\frac{\partial }{\partial \mathbf{r}_{2}}k_{2}^{(N)}(\mathbf{r}_{1},%
\mathbf{r}_{2},t)$ is given by the differential identity (\ref{App-X2})
reported in Appendix B. Thus, upon invoking the identity $\mathbf{n}%
_{23}\delta \left( \left\vert \mathbf{r}_{3}-\mathbf{r}_{2}\right\vert
-\sigma \right) =-\frac{\partial }{\partial \mathbf{r}_{3}}\overline{\Theta }%
\left( \left\vert \mathbf{r}_{3}-\mathbf{r}_{2}\right\vert -\sigma \right) $%
, one notices that an integration by parts can be performed also with
respect to $\mathbf{r}_{3}.$ This means that a procedure analogous to the
one used for the calculation of\textbf{\ }$\frac{\partial k_{1}^{(N)}(%
\mathbf{r}_{1},t)}{\partial \mathbf{r}_{1}}$ can be invoked and iterated at
all orders, i.e., up to the $(N-1)-$body occupation coefficient\ (see Eq.(%
\ref{App-X3}) in Appendix B). As a consequence the functional $K_{M}(\rho
_{1}^{(N)}(t),\mathbf{b})$ can be represented in terms of a finite sum of
the form $K_{M}(\rho _{1}^{(N)}(t),\mathbf{b})\equiv
\sum\limits_{j=1,N-1}K_{M}^{(j)}(\rho _{1}^{(N)}(\mathbf{x}_{1},t),\mathbf{b}%
)$ in which each term of the sum\textbf{\ }$K_{M}^{(j)}(\rho _{1}^{(N)}(%
\mathbf{x}_{1},t),\mathbf{b})$ is non-negative and symmetric.\ This implies
therefore that the same functional $K_{M}(\rho _{1}^{(N)}(t),\mathbf{b})$%
\textbf{\ }can be cast in the form
\begin{eqnarray}
&&K_{M}(\rho _{1}^{(N)}(t),\mathbf{b})=\int\limits_{\Gamma _{1(1)}}d\mathbf{x%
}_{1}\int\limits_{\Gamma _{1(2)}}d\mathbf{x}_{2}\overline{\Theta }%
_{1}^{(\partial \Omega )}(\overline{\mathbf{r}}_{1})\overline{\Theta }%
_{2}^{(\partial \Omega )}(\overline{\mathbf{r}}_{2})M(\mathbf{v}_{1},\mathbf{%
v}_{2},\mathbf{b})\times  \notag \\
&&F(\mathbf{r}_{1},\mathbf{r}_{2},t)\frac{\partial \widehat{\rho }_{1}^{(N)}(%
\mathbf{x}_{1},t)}{\partial \mathbf{r}_{1}}\cdot \frac{\partial \widehat{%
\rho }_{1}^{(N)}(\mathbf{x}_{2},t)}{\partial \mathbf{r}_{2}}\overline{\Theta
}\left( \left\vert \mathbf{r}_{2}-\mathbf{r}_{1}\right\vert -\sigma \right) ,
\label{REPRESENTATION OF KM}
\end{eqnarray}%
with $M(\mathbf{v}_{1},\mathbf{v}_{2},\mathbf{b})\geq 0$ being the total
directional kinetic energy (\ref{TOTAL DIRECTIONAL ENERGY}) and $F(\mathbf{r}%
_{1},\mathbf{r}_{2},t)$ a suitable real scalar kernel which is symmetric in
the variables $\mathbf{r}_{1}$ and $\mathbf{r}_{2}.$ Hence\textbf{\ }$%
K_{M}(\rho _{1}^{(N)}(t),\mathbf{b})$ actually defines a non-negative
functional. This proves the validity of the inequality (\ref{P3-1-3})
(Proposition \emph{P1}$_{1}$).

In a similar way also the remaining Propositions can be established. In
fact, invoking Eq.(\ref{MKI-6}) it follows that the inequalities (\ref%
{P3-1-2}) and (\ref{P3-1-3}) \textbf{- }and hence also Propositions \emph{P1}%
$_{2}$ and \emph{P1}$_{3}$ - manifestly hold too. Finally, regarding the
proof of Proposition\emph{\ P1}$_{4}$, one notices that $K_{M}(\rho
_{1}^{(N)}(t),\mathbf{b})\equiv 0$ if and only if identically $\frac{%
\partial }{\partial \mathbf{r}_{1}}\widehat{\rho }_{1}^{(N)}(\mathbf{x}%
_{1},t)\equiv 0$. Since $\widehat{\rho }_{1}^{(N)}(\mathbf{x}_{1},t)$ is by
construction a solution of the Master kinetic equation it follows that this
requires necessarily that $\widehat{\rho }_{1}^{(N)}(\mathbf{x}_{1},t)$ must
coincide with the local Maxwellian $\rho _{1M}^{(N)}(\mathbf{v}_{1})$ (see
Eq.(\ref{MAXWELLIAN-PDF})) and hence Eq.(\ref{P3-1-4}) must hold too under
the same realization (Proposition \emph{P1}$_{4}$). \textbf{Q.E.D.}

\bigskip

The conclusion is therefore that the definition of the MKI functional (\ref%
{MKI-functional-1}) given above in terms of $K_{M}(\rho _{1}^{(N)}(t),%
\mathbf{b})$ and $K_{Mo}$\textbf{\ }(see Eqs.(\ref{PRESCRIPTIONS FOR K_M and
K_Mo})) is indeed consistent with the physical prerequisites represented by
the MKI Prescriptions No.\#1 and No.\#2.

\subsection{2B - Proof of PMI for the Master kinetic equation}

The next step is to prove that the functional $I_{M}(\rho _{1o}^{(N)},%
\mathbf{b})$ defined above (see Eq.(\ref{MKI-6})) indeed exhibits a
monotonic time-decreasing behavior which is consistent with the MKI
Prescriptions No.\#3 and No.\#4, which are realized respectively by:

\begin{itemize}
\item the time derivative inequality (\ref{MKI-2}) and the conditions of
existence of kinetic equilibrium (\ref{MKI-4a});

\item the validity of the inequality $I_{M}(\rho _{1}^{(N)}(t))\leq 1$.
\end{itemize}

In order to reach the proofs of these properties let us preliminarily
determine\ the variation across a binary collision occurring between
particles $1$ and $2$\ of the total directional kinetic energy $M(\mathbf{v}%
_{1},\mathbf{v}_{2},\mathbf{b})$ (see Eq.(\ref{TOTAL DIRECTIONAL ENERGY})),
namely the phase-space scalar function $\Delta M(\mathbf{v}_{1},\mathbf{v}%
_{2},\mathbf{b})\equiv M(\mathbf{v}_{1}^{(+)},\mathbf{v}_{2}^{(+)},\mathbf{b}%
)-M(\mathbf{v}_{1}\mathbf{v}_{2},\mathbf{b})$. One obtains%
\begin{equation}
\Delta M(\mathbf{v}_{1},\mathbf{v}_{2},\mathbf{b})=\mathbf{b\cdot n}%
_{12}\left\vert \mathbf{n}_{12}\cdot \mathbf{v}_{12}^{(+)}\right\vert
\mathbf{v}_{12}^{(+)}\cdot \mathbf{b}-\left( \mathbf{b\cdot n}_{12}\right)
^{2}\left( \mathbf{n}_{12}\cdot \mathbf{v}_{12}^{(+)}\right) ^{2},
\label{directional eneergy VARIATION}
\end{equation}%
the rhs being expressed in terms of the outgoing particle velocities $(%
\mathbf{v}_{1}^{(+)},\mathbf{v}_{2}^{(+)})$ only. Then, the following
proposition holds.

\bigskip

\textbf{THM. 2 - Property of macroscopic irreversibility }(\textbf{Master
equation PMI theorem})

\emph{Let us assume that }$\rho _{1}^{(N)}(\mathbf{x}_{1},t)\equiv \rho
_{1}^{(N)}(\mathbf{r}_{1},\mathbf{v}_{1},t)$\emph{\ is an arbitrary
stochastic particular solution of the Master kinetic equation (\ref{App-1})
with initial condition\ }$\rho _{1o}^{(N)}(\mathbf{x}_{1})$ \emph{such that
the integral }$K_{M}(\rho _{1o}^{(N)}(\mathbf{x}_{1}),\mathbf{b})$\emph{\
exists and\ is non-vanishing. Then it follows that}

\begin{itemize}
\item \emph{Proposition P2}$_{1}$: \emph{one finds that for all }$t\geq
t_{o}:$%
\begin{eqnarray}
&&\left. \frac{\partial }{\partial t}K_{M}(\rho _{1}^{(N)}(t),\mathbf{b}%
)=-(N-1)\sigma ^{2}\int\limits_{U_{1(1)}}d\mathbf{v}_{1}\int%
\limits_{U_{1(2)}}d\mathbf{v}_{2}\int\limits_{\Omega }d\mathbf{r}_{1}%
\overline{\Theta }_{1}^{(\partial \Omega )}(\overline{\mathbf{r}}%
_{1})\int^{(-)}d\mathbf{\Sigma }_{21}\left\vert \mathbf{v}_{12}^{(+)}\cdot
\mathbf{n}_{12}\right\vert \right.  \notag \\
&&\left( \mathbf{b\cdot n}_{12}\right) ^{2}\left( \mathbf{n}_{12}\cdot
\mathbf{v}_{12}^{(+)}\right) ^{2}\frac{\partial \widehat{\rho }_{1}^{(N)}(%
\mathbf{r}_{1},\mathbf{v}_{1}^{(+)},t)}{\partial \mathbf{r}_{1}}\cdot \frac{%
\partial }{\partial \mathbf{r}_{2}}\widehat{\rho }_{1}^{(N)}(\mathbf{r}_{2}=%
\mathbf{r}_{1}+\sigma \mathbf{n}_{21},\mathbf{v}_{2}^{(+)}t)  \notag \\
&&\left. k_{2}^{(N)}(\mathbf{r}_{1},\mathbf{r}_{2}=\mathbf{r}_{1}+\sigma
\mathbf{n}_{21},t)\leq 0.\right.  \label{P2-2-2}
\end{eqnarray}

\item \emph{Proposition P2}$_{2}$: \emph{the inequality }%
\begin{equation}
\frac{\partial }{\partial t}I_{M}(\rho _{1}^{(N)}(t),\mathbf{b})\leq 0
\label{P2-2-1}
\end{equation}%
\emph{holds globally (i.e., identically for all }$t\geq t_{o}$\emph{)}%
\textbf{\ }\emph{so that necessarily\ }$I_{M}(\rho _{1}^{(N)}(t),\mathbf{b})$%
\emph{\ is globally defined too, being also prescribed so that}
\begin{equation}
I_{M}(\rho _{1}^{(N)}(t),\mathbf{b})\leq 1.  \label{P2-2-1bis}
\end{equation}

\item \emph{Proposition P2}$_{3}$: \emph{one finds that a given time }$t\in
I $ \emph{with} $t\geq t_{o}:$%
\begin{equation*}
\frac{\partial }{\partial t}K_{M}(\rho _{1}^{(N)}(t),\mathbf{b}%
)=0\Leftrightarrow \rho _{1}^{(N)}(\mathbf{x}_{1},t)\equiv \rho _{1M}^{(N)}(%
\mathbf{v}_{1}).
\end{equation*}%
\emph{Proof - }Consider first the proof of proposition \emph{P2}$_{1}$ which
requires evaluation of the partial time derivative\textbf{\ }$\frac{\partial
}{\partial t}K_{M}(\rho _{1}^{(N)}(t),\mathbf{b}).$ Upon invoking the first
form of the Master kinetic equation (see Eq.(\ref{App-0}) in Appendix A),
explicit differentiation of $K_{M}(\rho _{1}^{(N)}(t),\mathbf{b})$ delivers
\begin{eqnarray}
\frac{\partial }{\partial t}K_{M}(\rho _{1}^{(N)}(t),\mathbf{b})
&=&-\int\limits_{\Gamma _{1(1)}}d\mathbf{x}_{1}M(\mathbf{v}_{1},\mathbf{b})%
\overline{\Theta }_{1}^{(\partial \Omega )}(\overline{\mathbf{r}}%
_{1})k_{1}^{(N)}(\mathbf{r}_{1},t)\left( -\mathbf{v}_{1}\cdot \frac{\partial
}{\partial \mathbf{r}_{1}}\right) \frac{\partial ^{2}\widehat{\rho }%
_{1}^{(N)}(\mathbf{x}_{1},t)}{\partial \mathbf{r}_{1}\cdot \partial \mathbf{r%
}_{1}}  \notag \\
&&-\int\limits_{\Gamma _{1(1)}}d\mathbf{x}_{1}M(\mathbf{v}_{1},\mathbf{b})%
\overline{\Theta }_{1}^{(\partial \Omega )}\frac{\partial ^{2}\widehat{\rho }%
_{1}^{(N)}(\mathbf{x}_{1},t)}{\partial \mathbf{r}_{1}\cdot \partial \mathbf{r%
}_{1}}\left( \frac{\partial }{\partial t}\right) k_{1}^{(N)}(\mathbf{r}%
_{1},t),
\end{eqnarray}%
namely, upon integration by parts in the first integral on the rhs,%
\begin{equation}
\frac{\partial }{\partial t}K_{M}(\rho _{1}^{(N)}(t),\mathbf{b}%
)=-\int\limits_{\Gamma _{1(1)}}d\mathbf{x}_{1}M(\mathbf{v}_{1},\mathbf{b})%
\overline{\Theta }_{1}^{(\partial \Omega )}\frac{\partial ^{2}\widehat{\rho }%
_{1}^{(N)}(\mathbf{x}_{1},t)}{\partial \mathbf{r}_{1}\cdot \partial \mathbf{r%
}_{1}}\left( \frac{\partial }{\partial t}+\mathbf{v}_{1}\cdot \frac{\partial
}{\partial \mathbf{r}_{1}}\right) k_{1}^{(N)}(\mathbf{r}_{1},t).
\end{equation}%
Hence, thanks to the differential identity (\ref{DIFF-identity}) it follows:%
\begin{eqnarray}
&&\left. \frac{\partial }{\partial t}K_{M}(\rho _{1}^{(N)}(t),\mathbf{b}%
)=-(N-1)\int\limits_{\Gamma _{1(1)}}d\mathbf{x}_{1}M(\mathbf{v}_{1},\mathbf{b%
})\overline{\Theta }_{1}^{(\partial \Omega )}(\overline{\mathbf{r}}_{1})%
\frac{\partial ^{2}\widehat{\rho }_{1}^{(N)}(\mathbf{x}_{1},t)}{\partial
\mathbf{r}_{1}\cdot \partial \mathbf{r}_{1}}\int\limits_{\overline{\Gamma }%
_{1(2)}}d\mathbf{x}_{2}\mathbf{v}_{12}\cdot \mathbf{n}_{12}\times \right.
\notag \\
&&\delta (\left\vert \mathbf{r}_{1}-\mathbf{r}_{2}\right\vert -\sigma
)k_{2}^{(N)}(\mathbf{r}_{1},\mathbf{r}_{2},t)\widehat{\rho }_{1}^{(N)}(%
\mathbf{x}_{2},t).
\end{eqnarray}%
Performing an integration by parts with respect to $\mathbf{r}_{1}$ and upon
invoking the first differential identity (\ref{App-10a}) reported in
Appendix B one obtains therefore:%
\begin{equation}
\frac{\partial }{\partial t}K_{M}(\rho _{1}^{(N)}(t),\mathbf{b})=W_{M}(\rho
_{1}^{(N)}(t),\mathbf{b}),  \label{P2-2-2a}
\end{equation}%
\begin{eqnarray}
&&\left. W_{M}(\rho _{1}^{(N)}(t),\mathbf{b})\equiv \int\limits_{\Gamma
_{1(1)}}d\mathbf{x}_{1}M(\mathbf{v}_{1},\mathbf{b})\overline{\Theta }%
_{1}^{(\partial \Omega )}(\overline{\mathbf{r}}_{1})\frac{\partial \widehat{%
\rho }_{1}^{(N)}(\mathbf{x}_{1},t)}{\partial \mathbf{r}_{1}}(N-1)\right.
\notag \\
&&\times \int\limits_{\overline{\Gamma }_{1(2)}}d\mathbf{x}_{2}\mathbf{v}%
_{12}\cdot \mathbf{n}_{12}\frac{\partial }{\partial \mathbf{r}_{1}}\left[
\delta (\left\vert \mathbf{r}_{1}-\mathbf{r}_{2}\right\vert -\sigma )\right]
k_{2}^{(N)}(\mathbf{r}_{1},\mathbf{r}_{2},t)\widehat{\rho }_{1}^{(N)}(%
\mathbf{x}_{2},t),  \label{P2-2-2b}
\end{eqnarray}%
where $\frac{\partial }{\partial \mathbf{r}_{1}}\left[ \delta (\left\vert
\mathbf{r}_{1}-\mathbf{r}_{2}\right\vert -\sigma )\right] =-\frac{\partial }{%
\partial \mathbf{r}_{2}}\left[ \delta (\left\vert \mathbf{r}_{1}-\mathbf{r}%
_{2}\right\vert -\sigma )\right] $. Hence performing a further integration
by parts with respect to $\mathbf{r}_{2}$ and using the second differential
identity on Eq. (\ref{App-10a}) (see Appendix B) the previous equation
finally yields%
\begin{eqnarray}
&&\left. W_{M}(\rho _{1}^{(N)}(t),\mathbf{b})=(N-1)\int\limits_{\Gamma
_{1(1)}}d\mathbf{x}_{1}\overline{\Theta }_{1}^{(\partial \Omega )}(\overline{%
\mathbf{r}}_{1})\int\limits_{\overline{\Gamma }_{1(2)}}d\mathbf{x}_{2}%
\mathbf{v}_{12}\cdot \mathbf{n}_{12}M(\mathbf{v}_{1},\mathbf{v}_{2},\mathbf{b%
})\times \right.  \notag \\
&&\delta (\left\vert \mathbf{r}_{1}-\mathbf{r}_{2}\right\vert -\sigma
)k_{2}^{(N)}(\mathbf{r}_{1},\mathbf{r}_{2},t)\frac{\partial \widehat{\rho }%
_{1}^{(N)}(\mathbf{x}_{1},t)}{\partial \mathbf{r}_{1}}\cdot \frac{\partial
\widehat{\rho }_{1}^{(N)}(\mathbf{x}_{2},t)}{\partial \mathbf{r}_{2}},
\label{P2-CASO-gen}
\end{eqnarray}

where the symmetry property with respect to the exchange of states $\left(
\mathbf{x}_{1},\mathbf{x}_{2}\right) $ has been invoked. In the previous
equation the integration on the Dirac delta can be performed at once letting
\begin{eqnarray}
&&\left. \int\limits_{\Gamma _{1(1)}}d\mathbf{x}_{1}\overline{\Theta }%
_{1}^{(\partial \Omega )}\int\limits_{\overline{\Gamma }_{1(2)}}d\mathbf{x}%
_{2}\delta (\left\vert \mathbf{r}_{1}-\mathbf{r}_{2}\right\vert -\sigma
)=\sigma ^{2}\int\limits_{U_{1(1)}}d\mathbf{v}_{1}\int\limits_{U_{1(2)}}d%
\mathbf{v}_{2}\int\limits_{\Omega }d\mathbf{r}_{1}\overline{\Theta }%
_{1}^{(\partial \Omega )}(\overline{\mathbf{r}}_{1})\right.  \notag \\
&&\left[ \int^{(+)}d\mathbf{\Sigma }_{21}\left\vert \mathbf{v}_{12}\cdot
\mathbf{n}_{12}\right\vert -\int^{(-)}d\mathbf{\Sigma }_{21}\left\vert
\mathbf{v}_{12}\cdot \mathbf{n}_{12}\right\vert \right] ,
\end{eqnarray}%
where the solid-angle integrations in the two integrals on the rhs are
performed respectively on the outgoing $(+)$ and incoming $(-)$ particles.
Furthermore, it is obvious that thanks to the causal form of MCBC (see Eq.(%
\ref{bbb3}) in Appendix C) the integral on outgoing particles $\int^{(+)}d%
\mathbf{\Sigma }_{21}$ can be transformed to a corresponding integration on
incoming ones, namely $\int^{(-)}d\mathbf{\Sigma }_{21}.$ Thus, the
contributions in the two phase-space integrals only differ because of the
variation $\Delta M(\mathbf{v}_{1},\mathbf{v}_{2},\mathbf{b})$ of the total
directional kinetic energy of particles $1$ and $2.$ This implies that
\begin{eqnarray}
&&\left. W_{M}(\rho _{1}^{(N)}(t),\mathbf{b})=(N-1)\sigma
^{2}\int\limits_{U_{1(1)}}d\mathbf{v}_{1}\int\limits_{U_{1(2)}}d\mathbf{v}%
_{2}\int\limits_{\Omega }d\mathbf{r}_{1}\overline{\Theta }_{1}^{(\partial
\Omega )}(\overline{\mathbf{r}}_{1})\int^{(-)}d\mathbf{\Sigma }_{21}\times
\right.  \notag \\
&&\left\vert \mathbf{v}_{12}\cdot \mathbf{n}_{12}\right\vert \Delta M(%
\mathbf{v}_{1},\mathbf{v}_{2},\mathbf{b})\frac{\partial \widehat{\rho }%
_{1}^{(N)}(\mathbf{r}_{1},\mathbf{v}_{1}^{(+)},t)}{\partial \mathbf{r}_{1}}%
\cdot \frac{\partial \widehat{\rho }_{1}^{(N)}(\mathbf{r}_{2}=\mathbf{r}%
_{1}+\sigma \mathbf{n}_{21},\mathbf{v}_{2}^{(+)},t)}{\partial \mathbf{r}_{2}}%
\times  \notag \\
&&\left. k_{2}^{(N)}(\mathbf{r}_{1},\mathbf{r}_{2}=\mathbf{r}_{1}+\sigma
\mathbf{n}_{21},t),\right.  \label{P2-2-2c}
\end{eqnarray}%
where the solid-angle integration is performed on the incoming particles
whereas $\Delta M(\mathbf{v}_{1},\mathbf{v}_{2},\mathbf{b})$ is evaluated in
terms of the outgoing particles $(+)$ and therefore must be identified with
the second equation on the rhs of Eq.(\ref{directional eneergy VARIATION}).
Consider now the dependences in terms of the outgoing particle velocities $%
\mathbf{v}_{1}^{(+)}$ and $\mathbf{v}_{2}^{(+)}$ in the previous phase-space
integral. The velocity dependences contained in the factors $\left\vert
\mathbf{v}_{12}\cdot \mathbf{n}_{12}\right\vert $ and $\frac{\partial
\widehat{\rho }_{1}^{(N)}(\mathbf{r}_{1},\mathbf{v}_{1}^{(+)},t)}{\partial
\mathbf{r}_{1}}\cdot \frac{\partial \widehat{\rho }_{1}^{(N)}(\mathbf{r}_{2},%
\mathbf{v}_{2}^{(+)}t)}{\partial \mathbf{r}_{2}}$ are symmetric with respect
to the variables $\mathbf{v}_{1}^{(+)}$ and $\mathbf{v}_{2}^{(+)}.$ On the
other hand, as a whole, the same integral should remain unaffected with
respect to the exchange of the outgoing particle velocities $\mathbf{v}%
_{1}^{(+)}\Leftrightarrow \mathbf{v}_{2}^{(+)}.$ This means that the only
term in $\Delta M(\mathbf{v}_{1},\mathbf{v}_{2},\mathbf{b})$ which gives a\
(possibly) non-vanishing contribution is $-\left( \mathbf{b}\cdot \mathbf{n}%
_{12}\right) ^{2}\left( \mathbf{n}_{12}\cdot \mathbf{v}_{12}^{(+)}\right)
^{2}.$ As a consequence it is found that
\begin{eqnarray}
&&\left. \frac{\partial }{\partial t}K_{M}(\rho _{1}^{(N)}(t),\mathbf{b}%
)\equiv W_{M}(\rho _{1}^{(N)}(t),\mathbf{b})=-(N-1)\sigma
^{2}\int\limits_{U_{1(1)}}d\mathbf{v}_{1}\int\limits_{U_{1(2)}}d\mathbf{v}%
_{2}\times \right.  \notag \\
&&\int\limits_{\Omega }d\mathbf{r}_{1}\overline{\Theta }_{1}^{(\partial
\Omega )}(\overline{\mathbf{r}}_{1})\int^{(-)}d\mathbf{\Sigma }_{21}\frac{%
\partial \widehat{\rho }_{1}^{(N)}(\mathbf{r}_{1},\mathbf{v}_{1}^{(+)},t)}{%
\partial \mathbf{r}_{1}}\cdot \frac{\partial \widehat{\rho }_{1}^{(N)}(%
\mathbf{r}_{2}=\mathbf{r}_{1}+\sigma \mathbf{n}_{21},\mathbf{v}_{2}^{(+)},t)%
}{\partial \mathbf{r}_{2}}\times  \notag \\
&&\left. \left\vert \mathbf{v}_{12}^{(+)}\cdot \mathbf{n}_{12}\right\vert
\left( \mathbf{b\cdot n}_{12}\right) ^{2}\left( \mathbf{n}_{12}\cdot \mathbf{%
v}_{12}^{(+)}\right) ^{2}k_{2}^{(N)}(\mathbf{r}_{1},\mathbf{r}_{2}=\mathbf{r}%
_{1}+\sigma \mathbf{n}_{21},t)\leq 0,\right.  \label{P2-2-2d}
\end{eqnarray}%
and hence $\frac{\partial }{\partial t}K_{M}(\rho _{1}^{(N)}(t),\mathbf{b})$
is necessarily negative or null, the second case occurring only if $\frac{%
\partial \widehat{\rho }_{1}^{(N)}(\mathbf{r}_{1},\mathbf{v}_{1}^{(+)},t)}{%
\partial \mathbf{r}_{1}}\equiv 0$ and consequently\textbf{\ }$\frac{\partial
\widehat{\rho }_{1}^{(N)}(\mathbf{r}_{2}=\mathbf{r}_{1}+\sigma \mathbf{n}%
_{21},\mathbf{v}_{2}^{(+)}t)}{\partial \mathbf{r}_{2}}\equiv 0$ too.
\end{itemize}

The proof of Proposition P2$_{2}$ follows in a similar way. In fact, first,
one notices that thanks to the global validity of the $1-$body PDF \cite%
{noi7} the $1-$body PDF $\rho _{1}^{(N)}(t)$ necessarily belongs to the
functional class of stochastic PDFs $\left\{ \rho _{1}^{(N)}(\mathbf{x}%
_{1},t)\right\} $\textbf{\ }prescribed so that\emph{\ also the local
characteristic scale-length defined above}\textbf{\ }$L_{\rho }(t)$ (see Eq.
(\ref{b-lunghezza})) \emph{is larger than zero and finite}. As a consequence
it follows that both the functional $K_{M}(\rho _{1}^{(N)}(t),\mathbf{b})$
and $I_{M}(\rho _{1}^{(N)}(t),\mathbf{b})$ (see Eqs.(\ref{MKI-functional-1}))%
\textbf{\ }are globally defined too. Consider in fact the representation of $%
K_{M}(\rho _{1}^{(N)}(t),\mathbf{b})$ achieved in THM.1 and given by Eq.(\ref%
{RAPPRESENTAZIONE}). Next, let us notice that thanks to Eq.(\ref{b-lunghezza}%
) the characteristic scale length%
\begin{equation}
L_{\mu ,\min }\equiv \inf \left\{ \left\vert \frac{\partial \widehat{\rho }%
_{1}^{(N)}(\mathbf{x}_{1},t)}{\partial \mathbf{r}_{1}}\right\vert
^{-1}\right\}
\end{equation}%
is necessarily strictly positive. Then, upon noting that $k_{2}^{(N)}(%
\mathbf{r}_{1},\mathbf{r}_{2},t)\leq 1$ and $\int\limits_{\Gamma _{1(2)}}d%
\mathbf{x}_{2}\delta \left( \left\vert \mathbf{r}_{2}-\mathbf{r}%
_{1}\right\vert -\sigma \right) \widehat{\rho }_{1}^{(N)}(\mathbf{x}%
_{2},t)\leq \sup \left( \widehat{n}_{1}^{(N)}(\mathbf{r}_{2},t)\right)
<+\infty ,$ with $\widehat{n}_{1}^{(N)}(\mathbf{r}_{2},t)$ being the
velocity moment $\widehat{n}_{1}^{(N)}(\mathbf{r}_{2},t)=\int%
\limits_{U_{1(2)}}d\mathbf{v}_{2}\widehat{\rho }_{1}^{(N)}(\mathbf{r}_{2},%
\mathbf{v}_{2},t)$, it follows that%
\begin{equation}
K_{M}(\rho _{1}^{(N)}(t),\mathbf{b})\leq \frac{\left( N-1\right) }{L_{\mu
,\min }}\sup \left( \widehat{n}_{1}^{(N)}(\mathbf{r}_{2},t)\right)
\int\limits_{\Gamma _{1(1)}}d\mathbf{x}_{1}\overline{\Theta }_{1}^{(\partial
\Omega )}(\overline{\mathbf{r}}_{1})M(\mathbf{v}_{1},\mathbf{b})\widehat{%
\rho }_{1}^{(N)}(\mathbf{x}_{1},t),
\end{equation}%
where the integral on the rhs is necessarily bounded. This happens because $%
\rho _{1}^{(N)}(\mathbf{x}_{1},t)$ belongs to the functional class $\left\{
\rho _{1}^{(N)}(t)\right\} $ and therefore $\widehat{n}_{1}^{(N)}(\mathbf{r}%
_{2},t)$ is bounded, while, at the same time, the phase-space moments
indicated above necessarily exist. Furthermore, since\textbf{\ }$\frac{%
\partial }{\partial t}I_{M}(\rho _{1}^{(N)}(t),\mathbf{b})\equiv \frac{1}{%
K_{Mo}}\frac{\partial }{\partial t}K_{M}(\rho _{1}^{(N)}(t),\mathbf{b}),$
the inequality (\ref{P2-2-2d}) implies Eq.(\ref{P2-2-1}) and (\ref{P2-2-1bis}%
) too.\textbf{\ }Finally, since $\widehat{\rho }_{1}^{(N)}(\mathbf{r}_{1},%
\mathbf{v}_{1},t)$ is a solution of the Master kinetic equation $\frac{%
\partial }{\partial \mathbf{r}_{1}}\widehat{\rho }_{1}^{(N)}(\mathbf{x}%
_{1},t)\equiv 0$ occurs if and only if $\widehat{\rho }_{1}^{(N)}(\mathbf{x}%
_{1},t)$ coincides with a Maxwellian kinetic equilibrium of the type (\ref%
{MAXWELLIAN-PDF}). This\ result proves\ therefore also Proposition P2$_{3}.$
\textbf{Q.E.D.}

\bigskip

The implication of THM.2 is therefore that provided the initial value\emph{\
}$K_{M}(\rho _{1o}^{(N)}(\mathbf{x}_{1}),\mathbf{b})$ is non-vanishing then
necessarily:

\begin{itemize}
\item the functional $K_{M}(\rho _{1}^{(N)}(t),\mathbf{b})$ is monotonically
decreasing and thus $K_{M}(\rho _{1}^{(N)}(t),\mathbf{b})\leq K_{M}(\rho
_{1o}^{(N)}(\mathbf{x}_{1}),\mathbf{b})$;

\item similarly the MKI functional $I_{M}(\rho _{1}^{(N)}(t),\mathbf{b})$ is
monotonically decreasing too, i.e., $I_{M}(\rho _{1}^{(N)}(t),\mathbf{b}%
)\leq I_{M}(\rho _{1o}^{(N)}(\mathbf{x}_{1}),\mathbf{b})$;

\item both $K_{M}(\rho _{1}^{(N)}(t),\mathbf{b})$ and $I_{M}(\rho
_{1}^{(N)}(t),\mathbf{b})$ are non-negative.
\end{itemize}

\subsection{2C - Proof of the DKE property for the Master kinetic equation}

Let us now show that in validity of THMs. 1 and 2 the time-evolved $\rho
_{1}^{(N)}(\mathbf{x}_{1},t)$ necessarily must decay asymptotically for $%
\tau \equiv t-t_{o}\rightarrow +\infty $ to kinetic equilibrium, i.e., that
the limit function $\lim_{\tau \rightarrow +\infty }\rho _{1}^{(N)}(\mathbf{x%
}_{1},t)\equiv \rho _{1\infty }^{(N)}(\mathbf{x}_{1})$ exists and it
necessarily coincides with a Maxwellian kinetic equilibrium of the type (\ref%
{MAXWELLIAN-PDF}). In this regard the following proposition holds.

\bigskip

\textbf{THM. 3 - Asymptotic behavior of }$I_{M}(\rho _{1}^{(N)}(t),\mathbf{b}%
)$ (\textbf{Master equation-DKE theorem})

\emph{Let us assume that the initial condition} $\rho _{1o}^{(N)}(\mathbf{x}%
_{1})\in \left\{ \rho _{1o}^{(N)}(\mathbf{x}_{1})\right\} $\emph{\ is\ such
that the corresponding functional} $K_{M}(\rho _{1o}^{(N)}(\mathbf{x}_{1}),%
\mathbf{b})$ \emph{is non-vanishing, i.e., in view of THM.1 necessarily }$>0$%
\emph{. Then it follows that the corresponding time-evolved solution of the
Master kinetic equation} $\rho _{1}^{(N)}(\mathbf{x}_{1},t)$ \emph{in the
limit }$\tau \equiv t-t_{o}\rightarrow +\infty $ \emph{necessarily must
decay to kinetic equilibrium, i.e.,}
\begin{equation}
\lim_{\tau \rightarrow +\infty }\rho _{1}^{(N)}(\mathbf{x}_{1},t)=\rho
_{1M}^{(N)}(\mathbf{v}_{1}).  \label{P3-3-1}
\end{equation}%
\emph{Proof - }In order to reach the thesis it is sufficient to prove that
necessarily%
\begin{equation}
\lim_{\tau \rightarrow +\infty }\frac{\partial }{\partial t}I_{M}(\rho
_{1}^{(N)}(t),\mathbf{b})=0.  \label{EQQ.1}
\end{equation}%
In fact, let us assume "\textit{ad absurdum}" that $\frac{\partial }{%
\partial t}I_{M}(\rho _{1}^{(N)}(t),\mathbf{b})\leq -k^{2}$ with $k^{2}>0$
being a real constant. Then THM.2 (proposition P2$_{2}$) requires that%
\begin{equation}
\lim_{\tau \rightarrow +\infty }I_{M}(\rho _{1}^{(N)}(t),\mathbf{b})\leq
-\lim_{\tau \rightarrow +\infty }(t-t_{o})k^{2}=-\infty ,
\end{equation}%
a result which contradicts THM.1. This proves the validity of Eq.(\ref{EQQ.1}%
). Furthermore, by construction $\frac{\partial }{\partial t}I_{M}\equiv
\frac{1}{K_{Mo}}\frac{\partial }{\partial t}K_{M}$ and furthermore $\frac{%
\partial }{\partial t}K_{M}$ is identified with the functional $W_{M}(\rho
_{1}^{(N)}(t),\mathbf{b})\leq 0$ which is determined by Eq.(\ref{P2-2-2d}).
At this point one notices that, thanks to continuity of the functional $%
W_{M}(\rho _{1}^{(N)}(t),\mathbf{b}),$ the identity
\begin{equation}
\lim_{\tau \rightarrow +\infty }\frac{\partial }{\partial t}I_{M}(\rho
_{1}^{(N)}(\mathbf{x}_{1},t),\mathbf{b})=W_{M}(\rho _{1\infty }^{(N)}(%
\mathbf{x}_{1}),\mathbf{b})
\end{equation}%
holds, where, thanks to global existence of the $1-$body PDF (see Ref.\cite%
{noi7}), the limit function
\begin{equation}
\lim_{\tau \rightarrow +\infty }\rho _{1}^{(N)}(\mathbf{x}_{1},t)\equiv \rho
_{1\infty }^{(N)}(\mathbf{x}_{1})
\end{equation}%
necessarily exists. As a consequence Eq.(\ref{EQQ.1}) requires also the
equation%
\begin{equation}
W_{M}(\rho _{1\infty }^{(N)}(\mathbf{x}_{1}),\mathbf{b})=0
\end{equation}%
to hold. Upon invoking proposition P2$_{3}$ of THM.2 this implies that
necessarily $\rho _{1\infty }^{(N)}(\mathbf{x}_{1})=\rho _{1M}^{(N)}(\mathbf{%
v}_{1})$ so the thesis\ (\ref{P3-3-1}) is proved. Incidentally, thanks to
THM.1, this requires also that%
\begin{equation}
\lim_{\tau \rightarrow +\infty }I_{M}(\rho _{1}^{(N)}(t),\mathbf{b}%
)=I_{M}(\rho _{1\infty }^{(N)}(\mathbf{x}_{1}),\mathbf{b})=0.
\end{equation}

\textbf{Q.E.D}.

\subsection{2D - Remarks}

A few remarks are worth being pointed out regarding the results presented
above.

\begin{enumerate}
\item \emph{Remark \#1:} The choice of the MKI functional considered here
(see Eq.(\ref{MKI-functional-1})) is just one of the infinite particular
admissible realizations which meet the complete set of MKI-prescriptions
indicated above. In particular, the choice of the velocity moment $M(\mathbf{%
v}_{1},\mathbf{b})$ considered here (see Eq.(\ref{MOMENT-1})) remains in
principle arbitrary, since $\left\vert \mathbf{v}_{1}\cdot \mathbf{b}%
\right\vert ^{2}$ can be equivalently replaced, for example, by any factor
of the form $\left\vert \mathbf{v}_{1}\cdot \mathbf{b}\right\vert ^{2n},$
with $n\geq 1.$ Furthermore it is obvious that $M(\mathbf{v}_{1},\mathbf{b})$
can be replaced by any function of the form $M(\mathbf{v}_{1},\mathbf{b}%
)+\Delta M(\mathbf{v}_{1},\mathbf{b})$, being $\Delta M(\mathbf{v}_{1},%
\mathbf{b})$ prescribed in such a way that its contribution to $\frac{%
\partial }{\partial t}I_{M}$ vanishes identically so that the validity of
the inequality (\ref{P2-2-1}) in THM. 2 is preserved. This implies in turn
that the prescription of the MKI functional $I_{M}(\rho _{1}^{(N)}(t))$
remains in principle non-unique.

\item \emph{Remark \#2:} A possible issue is related to the requirement that
the renormalized $1-$body PDF, as the $1-$body PDF itself, are strictly
positive at all times and are non-vanishing. Here it is sufficient to state
that an elementary consequence of the theory of the Master kinetic equation
developed in Ref.\cite{noi3} is that, provided the corresponding initial $N-$%
body PDF set at a prescribed initial time $t_{o}$ is strictly positive in
the whole $N-$body phase-space, both the corresponding renormalized $1-$body
PDF, as the $1-$body PDF remain necessarily strictly positive globally in
time too and everywhere in the $1-$body phase-space.

\item \emph{Remark \#3:} It must be stressed that the signature of the time
derivative $\frac{\partial }{\partial t}I_{M}(\rho _{1}^{(N)}(t),\mathbf{b})$
actually depends crucially on the adoption of the causal form of MCBC (i.e.,
see Eq.(\ref{bbb1}) or (\ref{bbb3}) in Appendix C) rather than the
anti-causal one (given instead by Eq.(\ref{bbb2})). The first choice is
mandatory in view of the causality principle. Indeed, it is immediate to
prove that $\frac{\partial }{\partial t}I_{M}(\rho _{1}^{(N)}(t),\mathbf{b})$
changes signature if the anti-causal MCBC Eq.(\ref{bbb2}) is invoked.

\item \emph{Remark \#4:} THM.2 warrants that macroscopic irreversibility,
namely the inequality $\frac{\partial }{\partial t}I_{M}(\rho _{1}^{(N)}(t),%
\mathbf{b})\leq 0$ occurs specifically because of: a) the time-variation of
the $\mathbf{b}-$directional total kinetic energy which occurs at arbitrary
binary collision events; b) the occurrence of a velocity-space anisotropy in
the $1-$body PDF, i.e., the fact that the same PDF may not coincide with a
local Maxwellian PDF.

\item \emph{Remark \#5: }The existence of the limit function $%
\lim_{t\rightarrow +\infty }\rho _{1}^{(N)}(\mathbf{x}_{1},t)=\rho _{1\infty
}^{(N)}(\mathbf{x}_{1})$ follows uniquely as a consequence of the global
existence theorem holding for the Master kinetic equation \cite{noi7}.

\item \emph{Remark \#6: }Last but not least, the fact that the same limit
function may coincide or not with the Maxwellian kinetic equilibrium (\ref%
{MAXWELLIAN-PDF}) depends crucially on the functional setting prescribed for
the same PDF $\rho _{1}^{(N)}(\mathbf{x}_{1},t).$ More precisely, DKE can
only occur provided $\rho _{1}^{(N)}(\mathbf{x}_{1},t)$ is a suitably-smooth
stochastic PDF such that the MKI functional exists\ for the corresponding
initial PDF at time $t_{o},$ i.e., $\rho _{1}^{(N)}(\mathbf{x}%
_{1},t_{o})=\rho _{1o}^{(N)}(\mathbf{x}_{1}).$
\end{enumerate}

\bigskip

THMs 1-3 represent the main results reached in the paper of what may be
referred to as the \emph{PMI/DKE theory }for finite hard-sphere systems and%
\emph{\ }which have concerned the axiomatic formulation in such a context of
the notion of macroscopic irreversibility and the related one of decay to
kinetic equilibrium.

\section{3 - Consistency of MPI/DKE theory with microscopic dynamics}

The crucial problem which arises in the context of the ab initio-theory is
in some sense analogous to that occurring in the Boltzmann and Grad kinetic
theories. The question is in fact whether these phenomena are actually
consistent with the fundamental symmetry properties of the underlying
Boltzmann-Sinai CDS. The problem posed in the present section concerns, more
precisely, the consistency with the time-reversible, energy-conserving,
evolution of the underlying $N-$body Boltzmann-Sinai classical dynamical
system $S_{N}-$CDS.

\begin{enumerate}
\item \emph{First issue: consistency with the microscopic reversibility
principle} - This is related to the famous objection raised by Loschmidt to
the Boltzmann equation and Boltzmann H-theorem: i.e., \emph{whether} and
possibly also \emph{how} it may be possible to reconcile the validity of the
reversibility principle for the $S_{N}-$CDS with the manifestation of a
decay of the $1-$body PDF to kinetic equilibrium, i.e., the uniform
Maxwellian PDF of the form (\ref{MAXWELLIAN-PDF}), as predicted by the above
Master equation-DKE Theorem. That a satisfactory answer to this question is
actually possible follows from considerations which are based on the
axiomatic (ab initio) statistical description realized by the Master kinetic
equation.\ In this regard it is worth recalling the discussion reported
above concerning the role of MCBC regarding the functional $\frac{\partial }{%
\partial t}I_{M}(\rho _{1}^{(N)}(t)).$ In particular, it is obvious that the
signature depends on whether the causal (or anti-causal) form of MCBC is
invoked (see Appendix C). Such a choice is not arbitrary since, for
consistency with the causality principle, it must depend on the microscopic
arrow of time,\ i.e., the orientation of the time axis chosen for the
reference frame. Based on these premises, consistency between the occurrence
of macroscopic irreversibility associated with the DKE phenomenon and the
principle of microscopic reversibility can immediately be established.
Indeed, it is sufficient to notice that when a time-reversal or a
velocity-reversal is performed on the $S_{N}$ $-$CDS the form of the
collision boundary conditions (i.e., in the present case the MCBC provided
by Eq.(\ref{bbb1}) in Appendix C) must be changed, replacing them with the
corresponding anti-causal ones, i.e., Eq.(\ref{bbb2}). This implies that MKI
functional decreases in both cases, i.e., after performing the
time-reversal, so that no contradiction can possibly arise in this case
between THM.3 and the microscopic reversibility principle.

\item \emph{Second issue: consistency with Poincar\'{e} recurrence theorem -
}Similar considerations concern the consistency with the recurrence theorem
due to Poincar\'{e} as well as the conservation of total (kinetic) energy
for the $S_{N}$ $-$CDS. In fact, first, as it follows from Ref.\cite{noi3},
by construction the Master collision operator admits the customary Boltzmann
collisional invariants, including total kinetic energy of colliding
particles. Hence, total energy conservation is again warranted for $S_{N}$ $%
- $CDS. Second, regarding Poincar\'{e} recurrence theorem, it concerns the
Lagrangian phase-space trajectories of the $S_{N}$\ $-$CDS, i.e., the fact
that almost all of these trajectories return arbitrarily close - in a
suitable sense to be prescribed in terms of a distance defined on the $N-$%
body phase-space - to their initial condition after a suitably large
"recurrence time".\textbf{\ }Incidentally, its magnitude depends strongly
both on the same initial condition and the notion of distance to be
established on the same phase-space. Nevertheless, such a "recurrence
effect" influences only the Lagrangian time evolution of the $N-$body PDF
which occurs along the same Lagrangian $N-$body phase-space trajectories.
Instead, the same recurrence effect has manifestly no influence on the time
evolution of the Eulerian $1-$body PDF which is advanced in time in terms of
the Eulerian kinetic equation represented by the Master kinetic equation.
Therefore the mutual consistency of DKE and Poincar\'{e} recurrence theorem
remains obvious.
\end{enumerate}

Hence, in the framework of the axiomatic ab initio-theory based on the
Master kinetic equation the full consistency is warranted with the
microscopic dynamics of the underlying Boltzmann-Sinai CDS.

\section{4 - Physical implications}

Let us now investigate the physical interpretation and main implications
emerging\ from the PMI/DKE theory developed here. The first issue\ is
related to the physical mechanism at the basis of the PMI/DKE phenomenology.

It is well known that in the context of Boltzmann kinetic theory the
property of macroscopic irreversibility as well as the occurrence of the
DKE-phenomenon are both ascribed to the Boltzmann H-theorem, both in its
original formulation \cite{Boltzmann1972} and in its modified form
introduced by Boltzmann himself while attempting to reply \cite%
{Boltzmann1896} to Loschmidt objection \cite{Loschmidt1876} (see also Refs.
\cite{Cercignani1982,Lebowitz1993} together with different views on the
matter given in Refs. \cite{Droty2008,Uffink2015}). As recalled above, this
is expressed in terms of the production rate for the Boltzmann-Shannon
entropy $\frac{\partial }{\partial t}S(\rho _{1}(t)),$ with $S(\rho _{1}(t))$
being interpreted as a measure of the ignorance associated with a solution
of the Boltzmann equation.\ In fact the customary interpretation is that
they arise specifically because of the validity of the entropic inequality (%
\ref{Entropic enequality}), i.e., the monotonic increase of $S(\rho _{1}(t))$%
, and the corresponding entropic equality (\ref{entropic equality}) stating
a necessary and sufficient condition for kinetic equilibrium. Such a theorem
is actually intimately related with the equation itself. In fact both the
theorem and the equation generally hold only for stochastic PDFs $\rho
_{1}(t)=\rho _{1}(\mathbf{x}_{1},t)$ which are suitably-smooth and not for
distributions \cite{noi1}. According to Boltzmann's original interpretation,
however, both the Boltzmann equation and Boltzmann H-theorem should only
hold when the so-called Boltzmann-Grad limit is invoked, i.e. based\ on the
limit operator $L_{BG}\equiv \lim_{\substack{ N\rightarrow +\infty  \\ %
N\sigma ^{2}\sim O(1)}}$ (see Ref. \cite{noi11}).

In striking departure from such a picture:

\begin{itemize}
\item The axiomatic ab initio-theory based on the Master kinetic equation
and the present PMI/DKE theory are applicable to an arbitrary finite
Boltzmann-Sinai CDS. This means that they hold for hard-sphere systems
having a finite number of particles and with finite diameter and mass, i.e.,
without the need of invoking validity of asymptotic conditions.

\item The main departure with respect to Boltzmann kinetic theory arises
because, as earlier discovered (see in particular the related discussion
reported in Ref.\cite{noi11}),\ the Boltzmann-Shannon entropy associated
with an arbitrary stochastic $1-$body PDF $\rho _{1}^{(N)}(t)=\rho
_{1}^{(N)}(\mathbf{x}_{1},t)$ solution of the Master kinetic equation is
identically conserved. Thus both PMI and DKE are essentially unrelated to
the Boltzmann-Shannon entropy.

\item In the case of the Master kinetic equation the physical mechanism
responsible for the occurrence of both PMI and DKE is unrelated with the
Boltzmann-Shannon entropy. In fact, as shown here, it arises because of the
properties of the MKI functional $I_{M}(\rho _{1}^{(N)}(t),\mathbf{b})$ when
it is expressed in terms of an arbitrary stochastic PDF $\rho
_{1}^{(N)}(t)=\rho _{1}^{(N)}(\mathbf{x}_{1},t)$ solution of the Master
kinetic equation. The only requirement is that the initial PDF $\rho
_{1o}^{\left( N\right) }(\mathbf{x}_{1})$ is prescribed so that the
corresponding MKI functional $I_{M}(\rho _{1o}^{(N)}(\mathbf{x}_{1}),\mathbf{%
b})$ exists.

\item As shown here the MKI functional is a suitably-weighted phase-space
moment of $\rho _{1}^{(N)}(\mathbf{x}_{1},t)$ which can be interpreted as an
information measure for the same PDF, namely belongs to the interval $\left[
0,1\right] ,$ and exhibits a monotonic-decreasing time-dependence, i.e., the
property of macroscopic irreversibility.

\item In addition both $I_{M}(\rho _{1}^{(N)}(t),\mathbf{b})$ and its time
derivative $\frac{\partial }{\partial t}I_{M}(\rho _{1}^{(N)}(t),\mathbf{b})$
vanish identically if and only if the $1-$body PDF coincides with a
Maxwellian kinetic equilibrium of the type (\ref{MAXWELLIAN-PDF}). This
warrants in turn also the occurrence of the DKE-phenomenon for $\rho
_{1}^{(N)}(\mathbf{x}_{1},t)$, i.e., that for $t-t_{o}\rightarrow +\infty $
the same PDF must decay to a Maxwellian kinetic equilibrium of this type.

\item Finally, it is interesting to point out the peculiar behavior of the
MKI functional $I_{M}(\rho _{1}^{(N)}(t),\mathbf{b})$ and its time
derivative $\frac{\partial }{\partial t}I_{M}(\rho _{1}^{(N)}(t),\mathbf{b})$
when the Boltzmann-Grad limit is considered. In particular the $1-$ and $2-$%
body occupation coefficients $k_{1}^{(N)}(\mathbf{r}_{1},t)$ and $%
k_{2}^{(N)}(\mathbf{r}_{1},\mathbf{r}_{2},t)$ which appear in the Master
kinetic equation (see Appendix B, Eqs.(\ref{App-4}) and (\ref{App-5}))
become respectively%
\begin{equation}
\left\{
\begin{array}{c}
L_{BG}k_{1}^{(N)}(\mathbf{r}_{1},t)=1 \\
L_{BG}k_{2}^{(N)}(\mathbf{r}_{1},\mathbf{r}_{2},t)=1%
\end{array}%
.\right.
\end{equation}%
As a consequence the limit functionals $L_{BG}I_{M}(\rho _{1}^{(N)}(t),%
\mathbf{b})$ and $L_{BG}\frac{\partial }{\partial t}I_{M}(\rho _{1}^{(N)}(t),%
\mathbf{b}),$ are necessarily identically vanishing. This means that the
present theory applies properly when the exact Master kinetic equation is
considered and not to its asymptotic approximation obtained in the
Boltzmann-Grad limit, namely the Boltzmann kinetic equation (see Refs.\cite%
{noi3,noi7}).
\end{itemize}

An interesting issue, in the context of the PMI/DKE theory for the Master
kinetic equation, is the role of MCBC in giving rise to the phenomena of
macroscopic irreversibility and decay to kinetic equilibrium. Let us analyze
for this purpose the two cases represented by unary and binary hard-sphere
elastic collisions.

First, let us recall the customary treatment of collision boundary
conditions for unary collision events (also referred to as the so-called
mirror reflection CBC; see for example Cercignani \cite%
{Cercignani1969a,Cercignani1975}). This refers to the occurrence at a
collision time $t_{i}$\ of a single unary elastic collision for particle $1$%
\ at the boundary $\partial \Omega .$\textbf{\ }Let us denote by\textbf{\ }$%
\mathbf{n}_{1}$\textbf{\ }the inward normal to the stationary rigid boundary
$\partial \Omega $\ at the point of contact with the same particle and
respectively $\mathbf{x}_{1}^{(-)}(t_{1})=\left( \mathbf{r}_{1}(t_{1}),%
\mathbf{v}_{1}^{(-)}(t_{1})\right) $ and $\mathbf{x}_{1}^{(+)}(t_{1})=\left(
\mathbf{r}_{1}(t_{1}),\mathbf{v}_{1}^{(+)}(t_{1})\right) $ the incoming and
outgoing particle states while $\mathbf{v}_{1}^{(+)}$ is determined by the
elastic collision law for unary collisions, namely
\begin{equation}
\mathbf{v}_{1}^{(+)}=\mathbf{v}_{1}^{(-)}-2\mathbf{n}_{1}\mathbf{n}_{1}\cdot
\mathbf{v}_{1}^{(-)}.  \label{UNARY - elastic collision law}
\end{equation}%
Then, the PDF-conserving CBC for the $1-$body PDF requires that the
following identity holds%
\begin{equation}
\rho _{1}^{\left( N\right) }(\mathbf{x}_{1}^{(+)}(t_{1}),t_{i})=\rho
^{\left( N\right) }(\mathbf{x}_{1}^{(-)}(t_{i}),t_{i}),
\label{UNARY-PDF-CONSERVING CBC}
\end{equation}%
with $\rho _{1}^{\left( N\right) }(\mathbf{x}_{1}^{(+)}(t_{1}),t_{i})\equiv
\rho _{1}^{\left( N\right) (+)}(\mathbf{x}_{1}^{(+)}(t_{1}),t_{i})$ and $%
\rho ^{\left( N\right) }(\mathbf{x}_{1}^{(-)}(t_{i}),t_{i})\equiv \rho
^{\left( N\right) (-)}(\mathbf{x}_{1}^{(-)}(t_{i}),t_{i})$ denoting the
outgoing and incoming $1-$body PDF respectively. This identifies the
PDF-conserving CBC usually adopted in Boltzmann kinetic theory \cite%
{Boltzmann1972} (Grad \cite{Grad}; see also Refs.\cite{noi2,noi3}). The
obvious physical implication of Eq.(\ref{UNARY-PDF-CONSERVING CBC}) is that $%
\rho _{1}^{\left( N\right) }(\mathbf{x}_{1}^{(+)}(t_{1}),t_{i})$ (and $\rho
^{\left( N\right) }(\mathbf{x}_{1}^{(-)}(t_{i}),t_{i}))$ should be
necessarily an even function of the velocity component $\mathbf{n}_{1}\cdot
\mathbf{v}_{1}^{(-)}.$ Indeed as shown in Refs.\cite{noi2,noi3} the
PDF-conserving CBC (\ref{UNARY-PDF-CONSERVING CBC}) should be replaced with
a suitable CBC identified with the MCBC condition (see also Appendix C).
When realized in terms of its causal form (predicting the outgoing PDF in
terms of the incoming one) the MCBC for unary collisions is just:%
\begin{equation}
\rho _{1}^{\left( N\right) (+)}(\mathbf{x}_{1}^{(+)}(t_{1}),t_{i})=\rho
^{\left( N\right) (-)}(\mathbf{x}_{1}^{(+)}(t_{i}),t_{i}),
\end{equation}%
with $\rho ^{\left( N\right) (-)}(\mathbf{x}_{1}^{(+)}(t_{i}),t_{i})$
denoting the incoming $1-$body PDF evaluated in terms of the outgoing state $%
\mathbf{x}_{1}^{(+)}(t_{i}).$ Assuming left-continuity (see related
discussion in Ref.\cite{noi2}), this can then be identified with $\rho
^{\left( N\right) (-)}(\mathbf{x}_{1}^{(+)}(t_{i}),t_{i})\equiv \rho
^{\left( N\right) }(\mathbf{x}_{1}^{(+)}(t_{i}),t_{i}),$ thus yielding%
\begin{equation}
\rho _{1}^{\left( N\right) (+)}(\mathbf{x}_{1}^{(+)}(t_{1}),t_{i})=\rho
^{\left( N\right) }(\mathbf{x}_{1}^{(+)}(t_{i}),t_{i}).  \label{UNARY MCBC}
\end{equation}%
Eq.(\ref{UNARY MCBC}) provides the physical prescription for the collision
boundary condition, which is referred to as MCBC, holding for the $1-$body
PDF at arbitrary unary collision events. It is immediate to realize that the
function $\rho ^{\left( N\right) }(\mathbf{x}_{1}^{(+)}(t_{i}),t_{i})$ need
not generally be even with respect to the velocity component $\mathbf{n}%
_{1}\cdot \mathbf{v}_{1}^{(-)}.$ In addition Eq.(\ref{UNARY MCBC}), just as (%
\ref{UNARY-PDF-CONSERVING CBC}), also permits the existence of the customary
collisional invariants which in the case of unary collisions are $%
X=1,\left\vert \mathbf{n}_{1}\cdot \mathbf{v}_{1}^{(-)}\right\vert ,\mathbf{v%
}_{1}\cdot \left[ \underline{\underline{\mathbf{1}}}-\mathbf{n}_{1}\mathbf{n}%
_{1}\right] ,$ $v_{1}^{2}.$ As a consequence, one can show that Eq.(\ref%
{UNARY MCBC}) warrants at the same time also the validity of the so-called
no-slip boundary conditions for the fluid velocity field $\mathbf{V}(\mathbf{%
r}_{1},t)$ carried by the $1-$body PDF $\rho _{1}^{\left( N\right) }(\mathbf{%
x}_{1},t)$.

The treatment of MCBC holding for the $2-$body PDF in case of binary
collision events is analogous and is recalled for convenience in Eq.(\ref%
{bbb2}) of Appendix C.

Let us briefly analyze the qualitative physical implications of Eqs.(\ref%
{UNARY MCBC}) and (\ref{bbb2}) as far as the DKE theory is concerned. First,
we notice that unary collisions cannot produce in a proper sense a
velocity-isotropization effect since, as shown by Eq.(\ref{UNARY MCBC}), in
such a case MCBC gives rise only to a change in the velocity distribution
occurring during a unary collision due to a single component of the particle
velocity, namely $\mathbf{n}_{1}\cdot \mathbf{v}_{1}^{(-)}$.\textbf{\ }As a
consequence, this explains why unary collisions do not affect the rate of
change of the MKI functional (see THM.2). Second, Eq.(\ref{bbb2}) shows - on
the contrary - that binary collisions actually do affect by means of MCBC a
velocity-spreading for the $1-$\ and $2-$body PDF.\ In particular, since the
spreading effect occurs in principle for all components of
particle-velocities affecting both particles $1$\ and $2$, this explains why
binary collisions are actually responsible for the irreversible
time-evolution of the MKI functional (see THMs 2 and 3).

In turn, as implied by THM.3, DKE arises because of the phenomenon of
macroscopic irreversibility (THM.2). The latter arises due specifically to
the possible occurrence of a\textbf{\ }velocity-space anisotropy which
characterizes the $1-$body PDF when the same PDF differs locally from
kinetic equilibrium. In turn, this requires also that the $1-$body PDF
belongs to the functional class of admissible stochastic PDFs\ $\left\{ \rho
_{1}^{(N)}(\mathbf{x}_{1},t)\right\} $. In difference to Boltzmann kinetic
theory, however, the key physical role is actually ascribed to the MKI
functional $I_{M}(\rho _{1}^{(N)}(t))$ rather than the Boltzmann-Shannon
entropy $S_{1}(\rho _{1}^{(N)}(t))$. In fact, recalled above, the same
functional remains constant in time once the Master kinetic equation is
adopted. Rather, as shown by THM.2, it is actually the Master kinetic
information $I_{M}(\rho _{1}^{(N)}(t))$\ which exhibits the characteristic
signatures of macroscopic irreversibility.

The key differences arising between the two theories, i.e., the Boltzmann
equation-DKE and the Master equation-DKE, are of course related to the
different and peculiar intrinsic properties of the Boltzmann and Master
kinetic equations. In particular, as discussed at length elsewhere (see Refs.%
\cite{noi1,noi3}), precisely because the Boltzmann equation is only an
asymptotic approximation of the Master kinetic equation explains why a loss
of information occurs in Boltzmann kinetic theory and consequently the
related Boltzmann-Shannon entropy is not conserved.

The present investigation shows that in the context of the Master kinetic
equation, the macroscopic irreversibility property, i.e., the monotonic
time-decay behavior of the MKI functional, can be explained at a more
fundamental level, i.e., based specifically on the time-variation of the $%
\mathbf{b}-$directional total kinetic energy which occurs at arbitrary
binary collision events.

The Master equation-DKE theorem\ (THM.3) given above provides a
first-principle proof of the existence of the phenomenon of DKE occurring
for the kinetic description of a finite number of extended hard-spheres,
i.e., described by means of the Master kinetic equation. More precisely, the
DKE phenomenon affects the $1-$body PDFs belonging to the admissible
functional class $\left\{ \rho _{1}^{(N)}(\mathbf{x}_{1},t)\right\} $
determined according to THM.1 and requiring also that\ the local
characteristic scale-length $L_{\rho }(t)$\ associated with $\rho _{1}^{(N)}(%
\mathbf{x}_{1},t)$ is non-zero at all times.

\section{5 - Conclusions}

In this paper the problem of the property of microscopic irreversibility
(PMI) and decay to kinetic equilibrium (DKE) of the $1-$body PDF has been
addressed. In doing so original ideas and methods are adopted of the new ab
initio-theory for hard-sphere systems recently developed in the context of
Classical Statistical Mechanics \cite{noi1,noi2}.

These are not just small deviations from standard literature approaches.
Such developments, in fact, have opened up a host of exciting new subjects
of investigation and theoretical challenges in kinetic theory which arise
thanks to, or in the context of, the ab initio approach to kinetic theory.
Both are based in particular on the discovery of the Master kinetic equation
first reported in Ref. \cite{noi3}, equation which has been adopted also in
the present paper.

The ab initio-theory, and specifically the present paper, represent the
attempt to reach a new foundational basis and axiomatic physical description
of the classical statistical mechanics for hard-sphere systems. The topic
which has been pursued here - which represents also a challenging test of
the ab initio theory itself - concerns the investigation of the physical
origins of PMI\ and the related DKE phenomenon arising \emph{in finite }$N-$%
\emph{body hard-sphere systems.} These issues refer in particular to:

\begin{itemize}
\item \emph{The proof of the non-negativity of Master kinetic information
(THM.1, subsection 2A) together with the property of macroscopic
irreversibility (PMI; THM.2, subsection 2B).}

\item \emph{The establishment of THM.3 (subsection 2C) and the related proof
of the property of decay to kinetic equilibrium (DKE).}

\item \emph{The consistency of PMI and DKE with microscopic dynamics
(Section 3).}

\item \emph{The analysis of the main physical implications of DKE (Section
4).}
\end{itemize}

The theory presented here departs in several respects from previous
literature and notably from Boltzmann kinetic theory. The main differences
actually arise because of the non-asymptotic character of the new theory,
i.e., the fact that it applies to arbitrary dense or rarefied systems for
which the finite number and size of the constituent particles is accounted
for \cite{noi3}. In this paper basic consequences of the new theory have
been investigated which concern the phenomenon of decay to global kinetic
equilibrium.

The present results are believed to be crucial, besides in mathematical
research, for the physical applications of the ab initio-theory statistical
theory, i.e., the Master kinetic equation.\ Indeed, regarding challenging
future developments of the theory one should mention among others the
following examples of possible (and mutually-related) routes worth to be
explored. One is related to the investigation of the possible effects due to
arbitrarily prescribed, i.e., non-vanishing, initial (binary or multi-body)
\emph{phase-space} statistical correlations. As recalled above, in fact, the
Master equation is appropriate only when suitably-prescribed
configuration-space statistical correlations are taken into account. The
second goal concerns the investigation of the time-asymptotic properties of
the same kinetic equation, for which the present paper may represent a
useful basis. The third goal refers to the possible extension of the theory
to mixtures formed by hard spheres of different masses and diameter which
possibly undergo both elastic and anelastic collisions. \ Finally, the
fourth one concerns the investigation of hydrodynamic regimes for which a
key prerequisite is provided by the DKE theory established here.

\section{Acknowledgments}

This work is dedicated to the dearest memory of Flavia, wife of M.T.,
recently passed away. The investigation was developed in part within the
research projects: A) the Albert Einstein Center for Gravitation and
Astrophysics, Czech Science Foundation No. 14-37086G; B) the research
projects of the Czech Science Foundation GA\v{C}R grant No. 14-07753P; C)
the grant No. 02494/2013/RRC \textquotedblleft \textit{Kinetick\'{y} p\v{r}%
\'{\i}stup k proud\u{e}n\'{\i} tekutin}\textquotedblright\ (Kinetic approach
to fluid flow) in the framework of the \textquotedblleft Research and
Development Support in Moravian-Silesian Region\textquotedblright , Czech
Republic. Initial framework and motivations of the investigation were based
on the research projects developed by the Consortium for Magnetofluid
Dynamics (University of Trieste, Italy) and the MIUR (Italian Ministry for
Universities and Research) PRIN Research Program \textquotedblleft \textit{%
Problemi Matematici delle Teorie Cinetiche e Applicazioni}%
\textquotedblright\ (Mathematical Problems of Kinetic Theories and
Applications), University of Trieste, Italy. One of the authors (M.T.) is
grateful to the International Center for Theoretical Physics (Miramare,
Trieste, Italy) for the hospitality during the preparation of the manuscript.

\section{Appendix A: Realizations of the Master kinetic equation}

For completeness we recall here the two equivalent forms of the Master
kinetic equation \cite{noi3}. In terms of the renormalized $1-$body PDF $%
\widehat{\rho }_{1}^{(N)}(\mathbf{x}_{1},t)$ (see Eq.(\ref{App-00}) ) the
\emph{first form }of the same equation reads%
\begin{equation}
L_{1\left( 1\right) }\widehat{\rho }_{1}^{(N)}(\mathbf{x}_{1},t)=0,
\label{App-0}
\end{equation}%
with $L_{1\left( 1\right) }=\frac{\partial }{\partial t}+\mathbf{v}_{1}\cdot
\frac{\partial }{\partial \mathbf{r}_{1}}$ denoting the $1-$body
free-streaming operator. Hence it follows
\begin{equation}
L_{1\left( 1\right) }\rho _{1}^{(N)}(\mathbf{x}_{1},t)=\widehat{\rho }%
_{1}^{(N)}(\mathbf{x}_{1},t)L_{1\left( 1\right) }k_{1}^{(N)}(\mathbf{r}%
_{1},t),  \label{APP-0b}
\end{equation}%
where explicit evaluation of the rhs the last equation (see also Eq.(\ref%
{DIFF-identity}) below) yields
\begin{eqnarray}
&&\left. \widehat{\rho }_{1}^{(N)}(\mathbf{x}_{1},t)L_{1\left( 1\right)
}k_{1}^{(N)}(\mathbf{r}_{1},t)=\left( N-1\right) \sigma
^{2}\int\limits_{U_{1(2)}}d\mathbf{v}_{2}\int d\Sigma _{21}\mathbf{v}%
_{21}\cdot \mathbf{n}_{21}\right.  \notag \\
&&\overline{\Theta }^{\ast }(\mathbf{r}_{2})k_{2}^{(N)}(\mathbf{r}_{1},%
\mathbf{r}_{2},t)\widehat{\rho }_{1}^{(N)}(\mathbf{x}_{1},t)\widehat{\rho }%
_{1}^{(N)}(\mathbf{x}_{2},t),  \label{App-01}
\end{eqnarray}%
with $\overline{\Theta }^{\ast }(\mathbf{r}_{2})\equiv \overline{\Theta }%
_{i}^{(\partial \Omega )}(\overline{\mathbf{r}})$ and $k_{2}^{(N)}(\mathbf{r}%
_{1},\mathbf{r}_{2},t)$ being identified with the definitions given
respectively by Eqs.(\ref{App-2b}) and Eq.(\ref{App-4}) in Appendix B. Then,
consistent with Ref.\cite{noi3} and upon invoking the causal form of MCBC
(see Eq.(\ref{bbb3}) in Appendix C) the same equation can be written in the
equivalent \emph{second form} of the Master kinetic equation \cite{noi3}.
The corresponding initial-value problem, taking the form:%
\begin{equation}
\left\{
\begin{array}{c}
L_{1\left( 1\right) }\rho _{1}^{(N)}(\mathbf{x}_{1},t)-\mathcal{C}_{1}\left(
\rho _{1}^{(N)}|\rho _{1}^{(N)}\right) =0, \\
\rho _{1}^{(N)}(\mathbf{x}_{1},t_{o})=\rho _{1o}^{(N)}(\mathbf{x}_{1}),%
\end{array}%
\right.  \label{App-1}
\end{equation}%
\ can be shown to admit a unique global solution \cite{noi7}. Here the
notation is standard \cite{noi3}. Thus%
\begin{eqnarray}
&&\left. \mathcal{C}_{1}\left( \rho _{1}^{(N)}|\rho _{1}^{(N)}\right) \equiv
\left( N-1\right) \sigma ^{2}\int\limits_{U_{1(2)}}d\mathbf{v}%
_{2}\int^{(-)}d\Sigma _{21}\right.  \notag \\
&&\left[ \widehat{\rho }_{1}^{(N)}(\mathbf{r}_{1},\mathbf{v}_{1}^{(+)},t)%
\widehat{\rho }_{1}^{(N)}(\mathbf{r}_{2},\mathbf{v}_{2}^{(+)},t)-\widehat{%
\rho }_{1}^{(N)}(\mathbf{r}_{1},\mathbf{v}_{1},t)\widehat{\rho }_{1}^{(N)}(%
\mathbf{r}_{2},\mathbf{v}_{2},t)\right] \times  \notag \\
&&\left\vert \mathbf{v}_{21}\cdot \mathbf{n}_{21}\right\vert k_{2}^{(N)}(%
\mathbf{r}_{1},\mathbf{r}_{2},t)\overline{\Theta }^{\ast }(\mathbf{r}_{2})
\label{App-2}
\end{eqnarray}%
identifies the Master collision operator, while $\rho _{1o}^{(N)}(\mathbf{x}%
_{1})$ is the initial $1-$body PDF which belongs to the functional class $%
\left\{ \rho _{1o}^{(N)}(\mathbf{x}_{1})\right\} $ of stochastic, i.e.,
strictly-positive, smooth ordinary functions, $1-$body PDFs. Furthermore,
the solid-angle integral on the rhs of Eq.(\ref{App-2}) is now evaluated on
the subset in which $\mathbf{v}_{12}\cdot \mathbf{n}_{12}<0,$ while $\mathbf{%
r}_{2}$ identifies\emph{\ }$\mathbf{r}_{2}=\mathbf{r}_{1}+\sigma \mathbf{n}%
_{21}$, while $k_{1}^{(N)}(\mathbf{r}_{1},t)$ and $k_{2}^{(N)}(\mathbf{r}%
_{1},\mathbf{r}_{2},t)$ coincide respectively with the $1-$ and $2-$body
occupation coefficients \cite{noi3} and $\overline{\Theta }^{\ast }\equiv
\overline{\Theta }^{\ast }(\mathbf{r}_{i})$ is prescribed by$\ $%
\begin{equation}
\overline{\Theta }^{\ast }(\mathbf{r}_{i})\equiv \overline{\Theta }%
_{i}^{(\partial \Omega )}(\overline{\mathbf{r}})\equiv \overline{\Theta }%
\left( \left\vert \mathbf{r}_{i}-\frac{\sigma }{2}\mathbf{n}_{i}\right\vert -%
\frac{\sigma }{2}\right) ,  \label{App-2b}
\end{equation}%
with $\overline{\Theta }(x)$ being the strong Heaviside theta function $%
\overline{\Theta }(x)=\left\{
\begin{array}{lll}
1 &  & y>0 \\
0 &  & y\leq 0%
\end{array}%
\right. $.

Regarding the specific identification of the occupation coefficients
(recalled in Appendix B) let us preliminarily recall the notion of $S_{N}-$%
\emph{\ ensemble strong theta-function} $\overline{\Theta }^{(N)}.$ The
latter is prescribed, according to Ref.\cite{noi3}, by requiring that
\begin{equation}
\overline{\Theta }^{(N)}(\overline{\mathbf{r}})=1  \label{CONSTRAINT-1}
\end{equation}%
for all confguration vectors $\overline{\mathbf{r}}\equiv \left\{ \mathbf{r}%
_{1},...,\mathbf{r}_{N}\right\} $ belonging to the collisionless subset of $%
\Omega ^{(N)}$. This is identified with the open subset of the $N-$body
configuration domain $\Omega ^{(N)}\equiv \prod\limits_{i=1,N}\Omega $ in
which each of the particles of $S_{N}$ is not in mutual contact with any
other particle of $S_{N}$ or with the boundary $\partial \Omega $ of $\Omega
.$ This means that at any configuration $\overline{\mathbf{r}},$ $\overline{%
\Theta }^{(N)}(\overline{\mathbf{r}})$ can be prescribed as%
\begin{equation}
\overline{\Theta }^{(N)}(\overline{\mathbf{r}})\equiv \prod\limits_{i=1,N}%
\overline{\Theta }_{i}(\overline{\mathbf{r}})\overline{\Theta }%
_{i}^{(\partial \Omega )}(\overline{\mathbf{r}}).  \label{ENSEMBLE-THETA-1}
\end{equation}%
Here $\overline{\Theta }_{i}^{(\partial \Omega )}(\overline{\mathbf{r}})$
identifies the $i-$th particle "boundary" theta function%
\begin{equation}
\overline{\Theta }_{i}^{(\partial \Omega )}(\overline{\mathbf{r}})\equiv
\overline{\Theta }_{i}^{(\partial \Omega )}(\overline{\mathbf{r}}_{i})=%
\overline{\Theta }\left( \left\vert \mathbf{r}_{i}-\mathbf{r}%
_{Wi}\right\vert -\frac{\sigma }{2}\right) ,  \label{boundary theta function}
\end{equation}%
with $\mathbf{r}_{Wi}=\mathbf{r}_{i}-\rho \mathbf{n}_{i}$ and $\rho \mathbf{n%
}_{i}$ being the inward vector normal to the boundary belonging to the
center of the $i-$th particle having a distance $\rho /2$ from the same
boundary. Furthermore $\overline{\Theta }_{i}(\overline{\mathbf{r}})$ is the
"binary-collision" theta function. A possible identification of $\overline{%
\Theta }_{i}(\overline{\mathbf{r}})$ which warrants validity of Eq.(\ref%
{CONSTRAINT-1}) is given by the expression
\begin{equation}
\overline{\Theta }_{i}(\overline{\mathbf{r}})\equiv \prod\limits_{\substack{ %
j=1,N;  \\ i<j}}\overline{\Theta }\left( \left\vert \mathbf{r}_{i}-\mathbf{r}%
_{j}\right\vert -\sigma \right) ,  \label{ENSEMBLE.THETA-2A}
\end{equation}%
namely%
\begin{equation}
\left\{
\begin{array}{c}
\overline{\Theta }_{i}(\overline{\mathbf{r}})\equiv \prod\limits_{\substack{ %
j=1,N;  \\ i<j}}\overline{\Theta }_{ij}(\overline{\mathbf{r}}), \\
\overline{\Theta }_{ij}(\overline{\mathbf{r}})\equiv \overline{\Theta }%
\left( \left\vert \mathbf{r}_{i}-\mathbf{r}_{j}\right\vert -\sigma \right) .%
\end{array}%
\right.  \label{ENSEMBLE THETA-2A-FACTOR}
\end{equation}%
However, an equivalent possible prescription of $\overline{\Theta }_{i}(%
\overline{\mathbf{r}})$ is also provided by the alternative realization
obtained letting%
\begin{equation}
\left\{
\begin{array}{c}
\overline{\Theta }_{i}(\overline{\mathbf{r}})\equiv \prod\limits_{\substack{ %
j=1,N;  \\ i<j}}\prod\limits_{\substack{ m,n=1,N  \\ i<m<n}}\overline{\Theta
}_{ij}^{mn}(\overline{\mathbf{r}}), \\
\overline{\Theta }_{ij}^{mn}(\overline{\mathbf{r}})\equiv \overline{\Theta }%
\left( \left\vert \mathbf{r}_{i}-\mathbf{r}_{j}\right\vert -\sigma \right)
\times \\
\overline{\Theta }(\left\vert \mathbf{r}_{i}-\mathbf{r}_{m}\right\vert
+\left\vert \mathbf{r}_{i}-\mathbf{r}_{n}\right\vert -2\sigma )\overline{%
\Theta }(\left\vert \mathbf{r}_{m}-\mathbf{r}_{n}\right\vert +\left\vert
\mathbf{r}_{i}-\mathbf{r}_{m}\right\vert -2\sigma ).%
\end{array}%
\right.  \label{Factro-ENSEMBLE-THETA-2}
\end{equation}%
Indeed, in the subset of $\Omega ^{(N)}$ in which for all $i=1,N$ the rhs of
Eq.(\ref{ENSEMBLE.THETA-2A}) is identically equal to unity, the factor $%
\prod\limits_{\substack{ m,n=1,N  \\ i<m<n}}\overline{\Theta }(\left\vert
\mathbf{r}_{i}-\mathbf{r}_{m}\right\vert +\left\vert \mathbf{r}_{i}-\mathbf{r%
}_{n}\right\vert -2\sigma )\overline{\Theta }(\left\vert \mathbf{r}_{m}-%
\mathbf{r}_{n}\right\vert +\left\vert \mathbf{r}_{i}-\mathbf{r}%
_{m}\right\vert -2\sigma )$ is necessarily equal to unity too. Incidentally,
we notice in fact that the latter factor carries the contributions due to
triple collisions which are ruled out in the domain of validity of Eq.(\ref%
{CONSTRAINT-1}).

\section{Appendix B - Integral and differential identities for the
occupation coefficients}

One notices that although the definitions (\ref{Factro-ENSEMBLE-THETA-2})
and (\ref{ENSEMBLE.THETA-2A}) given in Appendix A for $\overline{\Theta }%
_{i}(\overline{\mathbf{r}})$ coincide in the collisionless subset of $\Omega
^{(N)},$ only the first one is applicable in the complementary collision
subset. Based on these premises in this appendix a number of integral and
differential identities holding for the $1-$ and $2-$body occupation
coefficients are displayed.

First, recalling Ref.\cite{noi3}, one notices that the realizations of the $%
1-$ and $s-$body occupation coefficients $k_{1}^{(N)}(\mathbf{r}%
_{i},t),k_{2}^{(N)}(\mathbf{r}_{1},\mathbf{r}_{2},t),$..., $k_{s}^{(N)}(%
\mathbf{r}_{1},\mathbf{r}_{2},..\mathbf{r}_{s},t)$ remain uniquely
prescribed by the $1-$body PDF, being given by%
\begin{eqnarray}
k_{1}^{(N)}(\mathbf{r}_{1},t) &\equiv &F_{1}\left\{ \prod\limits_{j=2,N}%
\frac{\rho _{1}^{(N)}(\mathbf{x}_{j},t)}{k_{1}^{(N)}(\mathbf{r}_{j},t)}%
\right\} ,  \label{App-4} \\
k_{2}^{(N)}(\mathbf{r}_{1},\mathbf{r}_{2},t) &\equiv &F_{2}\left\{
\prod\limits_{j=s+1,N}\frac{\rho _{1}^{(N)}(\mathbf{x}_{j},t)}{k_{1}^{(N)}(%
\mathbf{r}_{j},t)}\right\} ,  \label{App-5} \\
&&...  \notag \\
k_{s}^{(N)}(\mathbf{r}_{1},\mathbf{r}_{2},..\mathbf{r}_{s},t) &\equiv
&F_{s}\left\{ \prod\limits_{j=s+1,N}\frac{\rho _{1}^{(N)}(\mathbf{x}_{j},t)}{%
k_{1}^{(N)}(\mathbf{r}_{j},t)}\right\} ,  \label{App-6a}
\end{eqnarray}%
\textbf{\ }where $F_{s}$\ denotes the integral operator%
\begin{equation}
F_{s}\equiv \int\limits_{\Gamma _{N}}d\overline{\mathbf{x}}\overline{\Theta }%
^{(N)}(\overline{\mathbf{r}})\prod\limits_{i=1,s}\delta (\mathbf{x}_{i}-%
\overline{\mathbf{x}}_{i}).  \label{App-6}
\end{equation}%
Therefore, since in the collisionless subset of $\Omega ^{(N)}$ the
prescriptions (\ref{ENSEMBLE.THETA-2A}) and (\ref{Factro-ENSEMBLE-THETA-2})
are equivalent, in the same subset the $1-$\ and $2-$body occupation
coefficients, written in terms of Eq.(\ref{ENSEMBLE.THETA-2A}), become
explicitly%
\begin{eqnarray}
&&\left. k_{1}^{(N)}(\mathbf{r}_{1},t)=\int\limits_{\Gamma _{1(2)}}d\mathbf{x%
}_{2}\frac{\rho _{1}^{(N)}(\mathbf{x}_{2},t)}{k_{1}^{(N)}(\mathbf{r}_{2},t)}%
\Theta _{2}^{(\partial \Omega )}(\overline{\mathbf{r}})\overline{\Theta }%
\left( \left\vert \mathbf{r}_{2}-\mathbf{r}_{1}\right\vert -\sigma \right)
\times \right.  \notag \\
&&\int\limits_{\Gamma _{1(3)}}d\mathbf{x}_{3}\frac{\rho _{1}^{(N)}(\mathbf{x}%
_{3},t)}{k_{1}^{(N)}(\mathbf{r}_{3},t)}\Theta _{3}^{(\partial \Omega )}(%
\overline{\mathbf{r}})\prod\limits_{j=1,2}\overline{\Theta }\left(
\left\vert \mathbf{r}_{3}-\mathbf{r}_{j}\right\vert -\sigma \right) ....
\notag \\
&&\left. ...\int\limits_{\Gamma _{1(N)}}d\mathbf{x}_{N}\frac{\rho _{1}^{(N)}(%
\mathbf{x}_{N},t)}{k_{1}^{(N)}(\mathbf{r}_{N},t)}\Theta _{N}^{(\partial
\Omega )}(\overline{\mathbf{r}})\prod\limits_{j=1,N-1}\overline{\Theta }%
\left( \left\vert \mathbf{r}_{N}-\mathbf{r}_{j}\right\vert -\sigma \right)
,\right.  \label{App-7a}
\end{eqnarray}%
and%
\begin{eqnarray}
&&\left. k_{2}^{(N)}(\mathbf{r}_{1},\mathbf{r}_{2},t)=\int\limits_{\Gamma
_{1(3)}}d\mathbf{x}_{3}\frac{\rho _{1}^{(N)}(\mathbf{x}_{3},t)}{k_{1}^{(N)}(%
\mathbf{r}_{3},t)}\overline{\Theta }_{3}^{(\partial \Omega )}(\overline{%
\mathbf{r}})\prod\limits_{j=1,2}\overline{\Theta }\left( \left\vert \mathbf{r%
}_{3}-\mathbf{r}_{j}\right\vert -\sigma \right) \times \right.  \notag \\
&&\int\limits_{\Gamma _{1(4)}}d\mathbf{x}_{4}\frac{\rho _{1}^{(N)}(\mathbf{x}%
_{4},t)}{k_{1}^{(N)}(\mathbf{r}_{4},t)}\Theta _{4}^{(\partial \Omega )}(%
\overline{\mathbf{r}})\prod\limits_{j=1,3}\overline{\Theta }\left(
\left\vert \mathbf{r}_{4}-\mathbf{r}_{j}\right\vert -\sigma \right) ... \\
&&....\int\limits_{\Gamma _{1(N)}}d\mathbf{x}_{N}\frac{\rho _{1}^{(N)}(%
\mathbf{x}_{N},t)}{k_{1}^{(N)}(\mathbf{r}_{N},t)}\Theta _{N}^{(\partial
\Omega )}(\overline{\mathbf{r}})\prod\limits_{j=1,N-1}\overline{\Theta }%
\left( \left\vert \mathbf{r}_{N}-\mathbf{r}_{j}\right\vert -\sigma \right) .
\label{App-8a}
\end{eqnarray}%
Accordingly letting $\mathbf{n}_{jj}=\mathbf{r}_{uij}/\left\vert \mathbf{r}%
_{ij}\right\vert $ with $\mathbf{r}_{ij}=\mathbf{r}_{i}-\mathbf{r}_{j}$, one
notices that in the collisionless subset of $\Omega ^{(N)}$ the following
differential identities hold for all $s=1,N-1$:%
\begin{eqnarray}
&&\frac{\partial }{\partial \mathbf{r}_{1}}k_{1}^{(N)}(\mathbf{r}%
_{1},t)=\left( N-1\right) \int\limits_{\Gamma _{1(2)}}d\mathbf{x}_{2}\mathbf{%
n}_{12}\delta \left( \left\vert \mathbf{r}_{2}-\mathbf{r}_{1}\right\vert
-\sigma \right) \times  \notag \\
&&k_{2}^{(N)}(\mathbf{r}_{1},\mathbf{r}_{2},t)\overline{\Theta }%
_{2}^{(\partial \Omega )}(\overline{\mathbf{r}})\widehat{\rho }_{1}^{(N)}(%
\mathbf{x}_{2},t),  \label{App-X1} \\
&&\left\{
\begin{array}{c}
\frac{\partial }{\partial \mathbf{r}_{1}}k_{2}^{(N)}(\mathbf{r}_{1},\mathbf{r%
}_{2},t)=\left( N-2\right) \int\limits_{\Gamma _{1(3)}}d\mathbf{x}_{3}%
\mathbf{n}_{13}\delta \left( \left\vert \mathbf{r}_{3}-\mathbf{r}%
_{1}\right\vert -\sigma \right) \times \\
\prod\limits_{j=1,2;j\neq 1}\overline{\Theta }\left( \left\vert \mathbf{r}%
_{3}-\mathbf{r}_{j}\right\vert -\sigma \right) \overline{\Theta }%
_{3}^{(\partial \Omega )}(\overline{\mathbf{r}})k_{3}^{(N)}(\mathbf{r}_{1},%
\mathbf{r}_{2},\mathbf{r}_{3},t)\widehat{\rho }_{1}^{(N)}(\mathbf{x}_{3},t),
\\
\frac{\partial }{\partial \mathbf{r}_{2}}k_{2}^{(N)}(\mathbf{r}_{1},\mathbf{r%
}_{2},t)=\left( N-2\right) \int\limits_{\Gamma _{1(3)}}d\mathbf{x}_{3}%
\mathbf{n}_{23}\delta \left( \left\vert \mathbf{r}_{3}-\mathbf{r}%
_{2}\right\vert -\sigma \right) \times \\
\prod\limits_{j=1,2;j\neq 2}\overline{\Theta }\left( \left\vert \mathbf{r}%
_{3}-\mathbf{r}_{j}\right\vert -\sigma \right) \overline{\Theta }%
_{3}^{(\partial \Omega )}(\overline{\mathbf{r}})k_{3}^{(N)}(\mathbf{r}_{1},%
\mathbf{r}_{2},\mathbf{r}_{3},t)\widehat{\rho }_{1}^{(N)}(\mathbf{x}_{3},t),%
\end{array}%
\right.  \label{App-X2} \\
&&...............  \notag \\
&&\left\{
\begin{array}{c}
\frac{\partial }{\partial \mathbf{r}_{1}}k_{s}^{(N)}(\mathbf{r}_{1},\mathbf{r%
}_{2},..,\mathbf{r}_{s},t)=\left( N-s\right) \int\limits_{\Gamma _{1(s+1)}}d%
\mathbf{x}_{s+1}\mathbf{n}_{1s+1}\delta \left( \left\vert \mathbf{r}_{s+1}-%
\mathbf{r}_{1}\right\vert -\sigma \right) \times \\
\prod\limits_{j=1,s;j\neq 1}\overline{\Theta }\left( \left\vert \mathbf{r}%
_{s+1}-\mathbf{r}_{j}\right\vert -\sigma \right) \overline{\Theta }%
_{3}^{(\partial \Omega )}(\overline{\mathbf{r}})k_{s+1}^{(N)}(\mathbf{r}_{1},%
\mathbf{r}_{2},.,\mathbf{r}_{s+1},t)\widehat{\rho }_{1}^{(N)}(\mathbf{x}%
_{s+1},t) \\
\frac{\partial }{\partial \mathbf{r}_{2}}k_{s}^{(N)}(\mathbf{r}_{1},\mathbf{r%
}_{2},..,\mathbf{r}_{s},t)=\left( N-s\right) \int\limits_{\Gamma _{1(2)}}d%
\mathbf{x}_{s+1}\mathbf{n}_{2s+1}\delta \left( \left\vert \mathbf{r}_{s+1}-%
\mathbf{r}_{2}\right\vert -\sigma \right) \times \\
\prod\limits_{j=1,s;j\neq 2}\overline{\Theta }\left( \left\vert \mathbf{r}%
_{s+1}-\mathbf{r}_{j}\right\vert -\sigma \right) \overline{\Theta }%
_{3}^{(\partial \Omega )}(\overline{\mathbf{r}})k_{s+1}^{(N)}(\mathbf{r}_{1},%
\mathbf{r}_{2},.,\mathbf{r}_{s+1},t)\widehat{\rho }_{1}^{(N)}(\mathbf{x}%
_{s+1},t) \\
..... \\
\frac{\partial }{\partial \mathbf{r}_{s}}k_{s}^{(N)}(\mathbf{r}_{1},\mathbf{r%
}_{2},..,\mathbf{r}_{s},t)=\left( N-s\right) \int\limits_{\Gamma _{1(s+1)}}d%
\mathbf{x}_{s+1}\mathbf{n}_{ss+1}\delta \left( \left\vert \mathbf{r}_{s+1}-%
\mathbf{r}_{s}\right\vert -\sigma \right) \times \\
\prod\limits_{j=1,s;j\neq s}\overline{\Theta }\left( \left\vert \mathbf{r}%
_{s+1}-\mathbf{r}_{j}\right\vert -\sigma \right) \overline{\Theta }%
_{3}^{(\partial \Omega )}(\overline{\mathbf{r}})k_{s+1}^{(N)}(\mathbf{r}_{1},%
\mathbf{r}_{2},.,\mathbf{r}_{s+1},t)\widehat{\rho }_{1}^{(N)}(\mathbf{x}%
_{s+1},t)%
\end{array}%
\right.  \label{App-X3}
\end{eqnarray}%
As a consequence the following identities (the first one needed to evaluate
the rhs of Eq.(\ref{APP-0b}) in Appendix A)
\begin{eqnarray}
L_{1\left( 1\right) }k_{1}^{(N)}(\mathbf{r}_{1},t) &=&\left( N-1\right)
\int\limits_{\Gamma _{1(2)}}d\mathbf{x}_{2}\mathbf{v}_{21}\cdot \mathbf{n}%
_{21}\delta \left( \left\vert \mathbf{r}_{2}-\mathbf{r}_{1}\right\vert
-\sigma \right)  \notag \\
&&\overline{\Theta }^{\ast }(\mathbf{r}_{2})k_{2}^{(N)}(\mathbf{r}_{1},%
\mathbf{r}_{2},t)\widehat{\rho }_{1}^{(N)}(\mathbf{x}_{2},t),
\label{DIFF-identity}
\end{eqnarray}%
\begin{eqnarray}
&&\left. \frac{\partial ^{2}k_{1}^{(N)}(\mathbf{r}_{1},t)}{\partial \mathbf{r%
}_{1}\cdot \partial \mathbf{r}_{1}}=-\left( N-1\right) \int\limits_{\Gamma
_{1(2)}}d\mathbf{x}_{2}k_{2}^{(N)}(\mathbf{r}_{1},\mathbf{r}_{2},t)\delta
\left( \left\vert \mathbf{r}_{2}-\mathbf{r}_{1}\right\vert -\sigma \right)
\right.  \notag \\
&&\overline{\Theta }_{2}^{(\partial \Omega )}(\overline{\mathbf{r}})\mathbf{n%
}_{21}\cdot \frac{\partial }{\partial \mathbf{r}_{2}}\widehat{\rho }%
_{1}^{(N)}(\mathbf{x}_{2},t),  \label{Diff-identity-2}
\end{eqnarray}%
hold too. However, the alternative realization of the factor $\overline{%
\Theta }_{i}(\overline{\mathbf{r}})$ given by Eq.(\ref%
{Factro-ENSEMBLE-THETA-2}) (see Appendix A) has the virtue of excluding
explicitly multiple collisions. The consequence is that when such a
definition is adopted the differential identities%
\begin{equation}
\left\{
\begin{array}{c}
\delta \left( \left\vert \mathbf{r}_{2}-\mathbf{r}_{1}\right\vert -\sigma
\right) \frac{\partial }{\partial \mathbf{r}_{1}}k_{2}^{(N)}(\mathbf{r}_{1},%
\mathbf{r}_{2},t)=0, \\
\delta \left( \left\vert \mathbf{r}_{2}-\mathbf{r}_{1}\right\vert -\sigma
\right) \frac{\partial }{\partial \mathbf{r}_{2}}k_{2}^{(N)}(\mathbf{r}_{1},%
\mathbf{r}_{2},t)=0,%
\end{array}%
\right.  \label{App-10a}
\end{equation}%
both hold identically. The latter equations, in fact, manifestly hold also
in the collision subset where $\delta \left( \left\vert \mathbf{r}_{2}-%
\mathbf{r}_{1}\right\vert -\sigma \right) \neq 0.$

\section{Appendix C: Causal and anti-causal forms of collisional boundary
conditions}

For definiteness, let us denote respectively the outgoing and incoming $N-$%
body PDFs $\rho ^{(-)\left( N\right) }(\mathbf{x}^{\left( -\right)
}(t_{i}),t_{i})$ and $\rho ^{(+)\left( N\right) }(\mathbf{x}^{\left(
+\right) }(t_{i}),t_{i})$, with $\rho ^{(\pm )\left( N\right) }(\mathbf{x}%
^{\left( \pm \right) }(t_{i}),t_{i})=\lim_{t\rightarrow t_{i}^{\left( \pm
\right) }}\rho ^{\left( N\right) }(\mathbf{x}(t),t)$, where $\mathbf{x}%
^{(-)}(t_{i})$ and $\mathbf{x}^{(+)}(t_{i})$, with $\mathbf{x}^{\left( \pm
\right) }(t_{i})=\lim_{t\rightarrow t_{i}^{\left( \pm \right) }}\mathbf{x}%
(t),$ are the incoming and outgoing Lagrangian $N-$body states, their mutual
relationship being again determined by the collision laws holding for the $%
S_{N}-$CDS. Here it is understood that:

\begin{itemize}
\item The $S_{N}-$CDS is referred to a reference frame $O\left( \mathbf{r}%
,\tau \equiv t-t_{o}\right) $, having respectively spatial and time origins
at the point $O$ which belongs to the Euclidean space $%
\mathbb{R}
^{3}$ and at time $t_{o}\in I$.

\item In addition, by assumption the time-axis is oriented. Such an
orientation is referred to as \emph{microscopic arrow of time}.
\end{itemize}

For an arbitrary $N-$body PDF $\rho ^{\left( N\right) }(\mathbf{x},t)$
belonging to the extended functional setting and an arbitrary collision
event occurring at time $t_{i}$ two possible realizations of the MCBC can in
principle be given, both yielding a relationship between the PDFs $\rho
^{(+)\left( N\right) }$ and $\rho ^{(-)\left( N\right) }$. In the context of
the ab initio statistical approach\ based on the Master kinetic equation
\cite{noi1,noi2,noi3} these are provided by the two possible realizations of
the so-called \emph{modified CBC} (MCBC). When expressed in Lagrangian form
they are realized respectively either by the \emph{causal} and\emph{\
anti-causal MCBC}, namely%
\begin{equation}
\rho ^{(+)\left( N\right) }(\mathbf{x}^{\left( +\right) }(t_{i}),t_{i})=\rho
^{(-)\left( N\right) }(\mathbf{x}^{(+)}(t_{i}),t_{i}),  \label{bbb1}
\end{equation}%
or%
\begin{equation}
\rho ^{(-)\left( N\right) }(\mathbf{x}^{\left( -\right) }(t_{i}),t_{i})=\rho
^{(+)\left( N\right) }(\mathbf{x}^{(-)}(t_{i}),t_{i}).  \label{bbb2}
\end{equation}%
The corresponding Eulerian forms of the MCBC can easily be determined (see
Ref.\cite{noi2}). The one corresponding to Eq.(\ref{bbb1}) is, for example,
provided\ by the condition%
\begin{equation}
\rho ^{(+)\left( N\right) }(\mathbf{x}^{\left( +\right) },t)=\rho
^{(-)\left( N\right) }(\mathbf{x}^{(+)},t),  \label{bbb3}
\end{equation}%
where now $\mathbf{x}^{\left( +\right) }$ denotes again an arbitrary
outgoing collision state.

Once the time-axis is oriented, \textit{i.e.}, the \emph{microscopic} \emph{%
arrow of time} is prescribed, the validity of the causality principle in the
reference frame $\left( \mathbf{r},\tau \equiv t-t_{o}\right) $ manifestly
requires invoking Eq.(\ref{bbb1}). Indeed, Eq.(\ref{bbb1}) predicts the
future (\textit{i.e.}, outgoing) PDF from the past (incoming) one. Therefore
the choice (\ref{bbb1}) is the one which is manifestly consistent with the
causality principle. On the other hand, if the arrow of time is changed,
\textit{i.e.} the time-reversal transformation with respect to the initial
time (or time-origin) $t_{o},$ \textit{i.e.}, the map between the two
reference frames%
\begin{equation}
O\left( \mathbf{r},\tau \equiv t-t_{o}\right) \rightarrow O\left( \mathbf{r}%
,\tau ^{\prime }\right) ,
\end{equation}%
with $\tau ^{\prime }=-\tau $ is performed, it is obvious that for the
transformed\ reference frame $O\left( \mathbf{r},\tau ^{\prime }\right) $
the form of CBC consistent with causality principle becomes that given by
Eq.(\ref{bbb2}). Analogous conclusions hold if a velocity-reversal is
performed, implying the incoming states and corresponding PDF must be
exchanged with corresponding outgoing ones and vice versa.

\section{Appendix D: Treatment of case $N=2$}

For completeness let us briefly comment on the particular realization of
MPI/DKE theory which is achieved in the special case $N=2.$\ For this
purpose, one notices that - thanks to Eq.(\ref{App-5}) recalled in Appendix
B (see also Ref.\cite{noi3}) - in this case by construction $k_{2}^{(N)}(%
\mathbf{r}_{1},\mathbf{r}_{2},t)$\ simply reduces to%
\begin{equation}
k_{2}^{(N)}(\mathbf{r}_{1},\mathbf{r}_{2},t)\equiv 1.  \label{AppD-1}
\end{equation}%
Accordingly, once the same prescription is invoked, both the Master kinetic
equation (\ref{App-1}) and the corresponding Master collision operator (\ref%
{App-2}) remain formally unchanged. In a similar way it is important to
remark that the expression of the functional $\frac{\partial }{\partial t}%
K_{M}(\rho _{1}^{(N)}(t),\mathbf{b})\equiv W_{M}(\rho _{1}^{(N)}(t),\mathbf{b%
})$ given by Eq. (\ref{P2-2-2d}) is still correct also in such a case, being
now given by
\begin{eqnarray}
&&\left. \frac{\partial }{\partial t}K_{M}(\rho _{1}^{(N)}(t),\mathbf{b}%
)\equiv W_{M}(\rho _{1}^{(N)}(t),\mathbf{b})=-(N-1)\sigma
^{2}\int\limits_{U_{1(1)}}d\mathbf{v}_{1}\int\limits_{U_{1(2)}}d\mathbf{v}%
_{2}\times \right.  \notag \\
&&\int\limits_{\Omega }d\mathbf{r}_{1}\int^{(-)}d\mathbf{\Sigma }_{21}\frac{%
\partial \widehat{\rho }_{1}^{(N)}(\mathbf{r}_{1},\mathbf{v}_{1}^{(+)},t)}{%
\partial \mathbf{r}_{1}}\cdot \frac{\partial \widehat{\rho }_{1}^{(N)}(%
\mathbf{r}_{2}=\mathbf{r}_{1}+\sigma \mathbf{n}_{21},\mathbf{v}_{2}^{(+)}t)}{%
\partial \mathbf{r}_{2}}\times  \notag \\
&&\left. \left\vert \mathbf{v}_{12}^{(+)}\cdot \mathbf{n}_{12}\right\vert
\left( \mathbf{b\cdot n}_{12}\right) ^{2}\left( \mathbf{n}_{12}\cdot \mathbf{%
v}_{12}^{(+)}\right) ^{2}\leq 0.\right.  \label{AAPD-2}
\end{eqnarray}

It is then immediate to infer the validity of both the PMI theorem (THM.2)
and the DKE property for the Master kinetic equation (THM.3). As a
consequence one concludes that MPI/DKE theory holds also in the special case
$N=2$. This conclusion is not unexpected. In fact, binary collisions, as
indicated above, are responsible for the MPI/DKE phenomenology and in such a
case can only occur between particles $1$\ and $2$.


\begin{thebibliography}{99}
\bibitem{noi1} Massimo Tessarotto, Claudio Cremaschini and Marco Tessarotto,
Eur. Phys. J. Plus \textbf{128}, 32 (2013).

\bibitem{noi2} M. Tessarotto and C. Cremaschini, Phys. Lett. A \textbf{378},
1760 (2014).

\bibitem{noi3} M. Tessarotto and C. Cremaschini, Eur. Phys. J. Plus \textbf{%
129}, 157 (2014).

\bibitem{noi7} M. Tessarotto, C. Asci, C. Cremaschini, A. Soranzo and G.
Tironi, Eur. Phys. J. Plus \textbf{130}, 160 (2015).

\bibitem{noi8} M. Tessarotto, M. Mond and C. Asci, Eur. Phys. J. Plus
\textbf{132}, 213 (2017).

\bibitem{noi11} M. Tessarotto, C. Asci, C. Cremaschini, M. Mond, A. Soranzo
and G. Tironi, Found. Phys. \textbf{48} (3), 271-294 (2018).

\bibitem{Tessarotto1979} R. Clemente and M. Tessarotto, Trans.Th. Stat. Phys
\textbf{8}, 1 (1979).

\bibitem{Cercignani1969a} C. Cercignani, \textit{Mathematical methods in
kinetic theory}, Plenum Press, New York (1969).

\bibitem{Boltzmann1972} L. Boltzmann, \textit{Weitere Studien \"{u}ber das W%
\"{a}rmegleichgewicht unter Gasmolek\"{u}len}, Wiener Berichte, 66:
275--370; in WA I, paper 23 (1872).

\bibitem{Enskog} D. Enskog, Kungl. Svensk Vetenskps Akademiens \textbf{63},
4 (1921); (English translation by S. G. Brush).

\bibitem{CHAPMA-COWLING} S. Chapman and T. Cowling, \textit{The Mathematical
Theory of Nonuniform Gases}, Cambridge University Press (1951).

\bibitem{Grad} H. Grad, \textit{Thermodynamics of gases}, Handbook der
Physik \textbf{XII}, 205 (1958).

\bibitem{Cercignani1975} C. Cercignani, \textit{Theory and applications of
the Boltzmann equation}, Scottish Academic Press, Edinburgh and London
(1975).

\bibitem{Cercignani1988} C. Cercignani, \textit{The Boltzmann Equation and
Its Applications}, Springer Verlag (1988).

\bibitem{Sinai1970} Y.G. Sinai, Russ. Math. Surv. \textbf{25}, 137.(1970).

\bibitem{Sinai1989} Y.G. Sinai, \textit{Dynamical Systems II: Ergodic Theory
with Applications to Dynamical Systems and Statistical Mechanics}
(Springer-Verlag, Berlin, 1989).

\bibitem{Lanford1974} O.E. Lanford Jr. III, \textit{Time evolution of large
classical systems}, in Proc. Dynamical Systems, Theory and Applications,
1974 Battelle Rencontre on Dynamical Systems, Ed. J. Moser, Lecture Notes in
Physics, Vol.38 (Springer,Berlin, p.1 (1975).

\bibitem{Lanford1976} O.E. Lanford, Soc. Math. de France, Asterisque \textbf{%
40}, 117 (1976).

\bibitem{Lanford1981} O.E. Lanford, \textit{The hard sphere gas in the
Boltzmann-Grad limit,} Physica \textbf{106A}, 70 (1981).

\bibitem{Uffink2015} J. Uffink, G. Valente, Found Phys. \textbf{45},404
(2015).

\bibitem{Ardourel} V. Ardourel, Found Phys. \textbf{47,} 471 (2017).

\bibitem{Droty2008} A. Drory, S. Hist. Phil. Mod. Physics \textbf{39}, 889
(2008).

\bibitem{Loschmidt1876} J. Loschmidt, Akademie der Wissenschaften zu Wien
\textbf{73},128 (1876).

\bibitem{Boltzmann1896} L. Boltzmann, Wien. Ber.\textbf{\ 66}, 275 (1877);
in Boltzmann Vol.II, pp. 112--148 (1909).

\bibitem{Ehrenfest} P. Ehrenfest and T. Ehrenfest-Afanassjewa, \textit{The
Conceptual Foundations of the Statistical Approach in Mechanics,} Cornell
University Press, New York (1912).

\bibitem{Cercignani1982} C. Cercignani, Arch. Mech. Stosowanej\textbf{\ 34}%
(3), 231 (1982).

\bibitem{Lebowitz1993} J.L. Lebowitz, Phys. Today,\textbf{\ }November 1994,
p. 115 (1994).

\bibitem{Villani} C. Villani, \textit{Entropy production and convergence to
equilibrium for the Boltzmann equation}, 14th Int. Congress on Math. Physics
(28 July-2 August 2003 Lisbon, Portugal), Ed. J.C. Zambrini (University of
Lisbon, Portugal), Published by World Scientific Publishing Co. Pte. Ltd.,
ISBN \#9789812704016, pp. 130-144 (2006).

\bibitem{Gallavotti2014} G. Gallavotti, \textit{Nonequilibrium and
irreversibility}, Springer, Series "Theoretical and Mathematical Physics",
ISBN: 978-3-319-06757-5; arXiv e-print (arXiv:1311.6448) (2014).

\bibitem{van Beijeren} M.H. Ernst and H. van Beijeren, Physica \textbf{68},
437 (1973); \textbf{70},\textbf{\ }225 (1973); Phys. Lett. \textbf{43A}, 367
(1973).

\bibitem{ikt1} M. Tessarotto and M. Ellero, AIP Conf. Proc. \textbf{762},
108 (2005).

\bibitem{Crema2013a} C. Cremaschini and M. Tessarotto, Phys. Plasmas \textbf{%
20, }012901 (2013).

\bibitem{Crema2013b} C. Cremaschini, Z. Stuchlik and M. Tessarotto, Physics
of Plasmas\textbf{\ 20}, 052905 (2013).

\bibitem{Crema2014} C. Cremaschini, M. Tessarotto and Z. Stuchlik, Physics
of Plasmas\textbf{\ 21}, 032902 (2014).
\end{thebibliography}
\end{document}